\documentclass[aps,pre,twocolumn,superscriptaddress,floatfix]{revtex4-2}

\usepackage{ifpdf}
\usepackage{bm}
\usepackage{latexsym}
\usepackage{color}
\usepackage{pstricks}
\usepackage{figsize}
\usepackage{color}
\usepackage{graphicx}
\usepackage{float}
\usepackage{color,soul}
\usepackage{amssymb}
\usepackage{hyperref}
\usepackage{amsmath}
\usepackage{psfrag}
\usepackage{verbatim}
\usepackage{bbold}
\usepackage[titletoc,title]{appendix}
\usepackage{listings} 
\usepackage{textcomp} 
\usepackage{enumitem}
\setlist{noitemsep,leftmargin=*}

\hypersetup{
colorlinks=true,
urlcolor= blue,
citecolor=blue,
linkcolor= blue,
bookmarks=true,
bookmarksopen=false,
}

\usepackage{ifthen}
\usepackage{ifpdf}

\usepackage{latexsym}
\usepackage{amsmath}
\usepackage{amssymb}
\usepackage{bm}
\usepackage{wasysym}

\ifpdf
\usepackage{graphicx}
\usepackage{epstopdf}
\else
\usepackage{graphicx}
\usepackage{epsfig}
\fi
\DeclareMathOperator{\Tr}{Tr}
\DeclareMathOperator{\la}{\langle}
\DeclareMathOperator{\ra}{\rangle}

\newcommand{\p}{\partial}

\newcommand{\hide}[1]{}

\begin{document}

\title{The polaronic effect of a metal layer on variable range hopping}

\author{Ofek Asban}
\affiliation{Department of Physics, Ben-Gurion University of the Negev, Beer-Sheva 84105, Israel} 
\author{Alexander Burin}
\affiliation{Department of Chemistry, Tulane University, New Orleans, Louisiana 70118, USA}
\author{Alexander Shnirman}
\affiliation{Institut f\"ur Theorie der Kondensierten Materie and Institut f\"ur Quantenmaterialien und Technologien, Karlsruhe Institute of Technology, Karlsruhe, Germany}
\author{Moshe Schechter}
\affiliation{Department of Physics, Ben-Gurion University of the Negev, Beer-Sheva 84105, Israel}

\begin{abstract}
We investigate the effect of a metal plate on the variable range hopping (VRH) conductivity of a two dimensional electron-glass (EG) system. The VRH conductivity is known to have a stretched exponential dependence on temperature, with an exponent $p$ that depends on the shape of the EG's single particle density of states (DOS). For constant DOS $p=1/3$ and for linear DOS $p=1/2$, also known as Mott's and Efros-Shklovskii's VRH respectively. The presence of the plate causes two effects on the EG system, static and dynamic. The well known static effect accounts for the additional screening of the Coulomb repulsion in the EG and for the partial filling of the Coulomb gap in the DOS. This in turn causes an increase of the conductivity at very low temperatures. Here we investigate the complementary dynamical effect, which is related to the polaronic phenomena. Our main result is the dynamical suppression of the standard phonon assisted hopping and, thus, suppression 
of the conductivity in a much wider range of temperatures as compared to the low temperature static effect.
The relation to experiments is discussed.
\end{abstract}

\maketitle

\section{Introduction}
\label{Introduction}

Disordered systems have attracted much attention since Anderson's seminal work on localization transition~\cite{PhysRev.109.1492}. 
Deep in the localized phase transport is dominated by phonon assisted hopping.
The characteristic behavior is the well known variable rang hopping (VRH) with the conductivity showing 
the stretched exponential dependence on temperature:
\begin{equation}
\label{VRH-form}
\sigma \propto e^{-(T_0/T)^p}.
\end{equation}
Neglecting electron-electron interaction Mott~\cite{N.F.Mott} obtained the above expression for the conductivity with $T_0=(g_0 \xi^D)^{-1}$, where $\xi$ is the localization length, $g_0$ is the constant single particle density of states (DOS) in vicinity of the Fermi energy, $D$ is the dimension of the system, and $p=1/(D+1)$. Further, considering the Coulomb interactions Efros and Shklovskii~\cite{Efros_1975} (ES) found that the DOS has a soft gap (Coulomb gap) around the Fermi energy in the form of $g(E)\propto |E|^{D-1}$ and obtained $p=1/2$ for all dimensions, with $T_0 = e^2/\kappa\xi$, where $\kappa$ is the dielectric constant in natural units. Furthermore, a crossover as a function of decreasing temperature from Mott's to ES VRH conductivity was found theoretically and experimentally~\cite{PhysRevB.40.1216,PhysRevB.44.3599,Sarachik_2002,PhysRevB.80.245214}. The crossover is caused mostly by the formation of the Coulomb gap.

The effect of long-range interactions on the VRH conductivity has been further investigated by placing a metal layer in proximity to the disordered sample~\cite{PhysRevLett.56.643,HU199565,VANKEULS1996945,Adkins_1998,EStoMott3,PhysRevB.99.184201}, and similarly in quantum dot arrays~\cite{EStoMottQD1,EStoMottQD2}. The metallic plate is separated from the disordered sample by an insulating layer, whose thickness, $d$, is usually of the order of the typical nearest hopping distance in the disordered sample. The main effect of the metallic plate is considered to be enhanced screening in the disordered sample. For large distances ($r \gg d$) the screened interactions acquire a dipole form ($\sim 1/r^3$), giving rise to an approximately constant DOS at the center of the Coulomb gap~\cite{PhysRevB.49.13721,PhysRevB.99.184201}. Under these circumstances one should expect a reentrance of the Mott regime and an enhanced conductivity at low temperatures, which was indeed predicted theoretically~\cite{PhysRevB.49.13721} and measured experimentally~\cite{EStoMottQD1,EStoMottQD2,EStoMott3}. 
We denote this effect as {\it static effect}.

Yet, other experiments~\cite{PhysRevLett.56.643, Adkins_1998} in different materials show an opposite effect, where in the temperature range available for the experiment the metal plate induces: (1) an overall reduction in the conductivity and (2) 
activation ($p=1$) functional dependence of the conductivity at lowest available temperatures. An explanation for the activation behaviour in certain temperature regimes was provided by Larkin and Khmelnitskii~\cite{larkin1982activation} by an accurate account for the length-dependent screening.

In this paper we investigate the complementary {\it dynamical} effect of the electrons in the metallic layer on the VRH conductivity in the EG layer. This is a polaronic effect related to the dynamical rearrangement of electrons in the metallic layer resulted from the hopping of an electron in EG. 

The essence of this effect can easily be understood in a hypothetical situation of a metallic plate 
being kept at zero temperature whereas the EG and the phonons have a finite temperature. 
To each charge configuration of the EG there corresponds a ground state of the electrons in the 
metallic plate. These states are mostly orthogonal to each other. Their energies are fully accounted in 
the effect of static screening. Directly after a hopping event of an electron in the EG, the electrons in the metallic plate are no longer in their ground state (the new ground state is orthogonal to the old one). Thus extra energy has to be supplied by the thermal phonons to the electrons in the metallic plate on top of the activation energy provided to the hopping electron in the EG. This reduces the conductivity in the EG. We show that this effect dominates even if the metallic plate has the same temperature as the EG and the phonons.

We describe the system as an electron-glass coupled to two uncorrelated environments: phonons, and electrons in the metallic layer. The phonons are responsible for the original VRH mechanism (within the single-phonon approximation), and the electrons in the metallic plate both {\it statically} screen the Coulomb interaction and {\it dynamically} dress the tunneling amplitude. In principle the electronic environment could also provide the activation energy for VRH, yet this mechanism turns out to be 
subdominant.  

We use a field theoretical approach in order to obtain an effective action for the EG. 
We extract the part of the effective action responsible for the static screening and combine it with the original unscreened action of the EG. We then derive the conductivity to leading order in the dynamically dressed tunneling amplitude. We find that the polaronic effect has an approximated logarithmic dependence on hopping distance (for $r>d$ where $r$ is the hopping distance) and, therefore, it practically does not change the exponent $p$. This is in contrast to the static effect as discussed above. Yet, the polaronic dressing suppresses the tunneling amplitude and thus the conductivity. We find a wide temperature regime where the polaronic effect is dominant compared to the static effect, resulting in an overall reduction of the conductivity.

The paper is organized as follows. In Sec~\ref{MainResults} we present the main ideas and the results of the paper. In Sec.~\ref{ModelActionSec} we derive the effective action that consists of the EG model coupled to bosonic
field that represents both the phonon displacement field in the EG system and the potential field of the metal. By solving the saddle-point equations we show how the EG-metal interaction is screened. This is crucial for setting the right scale of the effective interaction between the EG system and the electronic bath. In Sec.~\ref{ConductivitySec} we obtain a general expression for the conductance between two localized states in the EG system. In Sec.~\ref{WeakDisorder} we further consider a more realistic scenario of a diffusive metallic plate. Finally, in Sec.~\ref{MacroCond} we present a regime where the polaronic effect is dominant as compared to the static effect and demonstrate the reduction in conductivity as a function of temperature. We then discuss our results in view of experimental data. Finally, we conclude in Sec.~\ref{Conclusions}.

\section{Main Results}
\label{MainResults}

We first review here our main results. The technical details are given in the following chapters. The physical picture described in Sec.~\ref{Introduction} is fully contained is the conductance 
$\sigma_{ij}$ between two localized EG sites $i$ and $j$, which is needed to evaluate the conductivity of the EG within the resistor network model. The conductance takes the form
\begin{equation}
\label{Conductance}
\sigma_{ij} = 2\pi\beta |t_{i j}|^2 n_i(1-n_j) P(E_{ij},r_{ij})\, .
\end{equation}
Here $\beta$ is the inverse temperature, $t_{ij}$ is the tunneling amplitude between sites $i$ and $j$, $n_i$ is the Fermi occupation on site $i$ with energy $E_i$, $E_{ij} = E_i-E_j$, $r_{ij} = r_i-r_j$ is the distance between sites $i$ and $j$ and $P(E_{ij},r_{ij})$ is the probability per energy for the EG system to emit (absorb) energy $E_{ij}$ to (from) the phononic and electronic environments for $E_{ij} > 0$ ($E_{ij} < 0$). The static screening by the metallic plate is already taken into account in the energies $E_i$ and, most importantly, in their density of states. The $r_{ij}$ dependence originates from the interaction of the extended phononic and electronic environments with the localized EG. The $r_{ij}$ dependence is crucial for the polaronic influence on the VRH hopping as further explained in Sec.~\ref{ConductivitySec}. The function $P(E_{ij},r_{ij})$ is given by a convolution of the contributions 
of the two environments:
\begin{equation}
\label{Convolution}
P(E,r) = \int_{-\infty}^{\infty} dE' P^{el}(E',r)P^{ph}(E-E',r)\, ,
\end{equation}
where $P^{el}$ and $P^{ph}$ represent the electronic and phononic environments respectively, the energies and distances are represented in a continuous form, $r_{ij}\rightarrow r$ and $E_{ij} \rightarrow E = E_I-E_F$. Eq.~(\ref{Convolution}) emphasizes the distribution of the energy emitted (absorbed) by the EG between the two environments. The main effect discussed in this paper 
relates to the regime in which the activated tunneling takes place, i.e., $E<0$, $|E|>T$. 
Due to its Ohmic spectrum the distribution $P^{el}$ is concentrated at low positive values (with respect to the cutoff frequency of the electronic environment) of $E'$ [see Eq.~(\ref{PofE-EH-Ohmic})]. 
This "forces" the phonons to provide extra activation energy to the electron-hole (e-h) excitations in the metal (see discussion in Appendix~\ref{MicroscopicOrigin}). This requires phonons of higher frequency, whose thermal occupation is smaller, which results in a lower conductivity.

The polaronic effect results in a total reduction of the conductivity obtained in Eq.~(\ref{CondSinglePh}) and Eq.~(\ref{PofE-EH-Ohmic}) and plotted in Fig.~\ref{PolaronReduction}. The apparent weaker temperature dependence than found experimentally is mainly a consequence of the strong screening in the metal. In Fig.~\ref{PolaronReductionFit} we show that by assuming smaller screening and thus larger effective interaction between the EG electrons the metal electrons, a good fit to experiment is obtained.

\section{The model and derivation of the effective action}
\label{ModelActionSec}

\subsection{The model}

\begin{figure}
\includegraphics[scale=0.4]{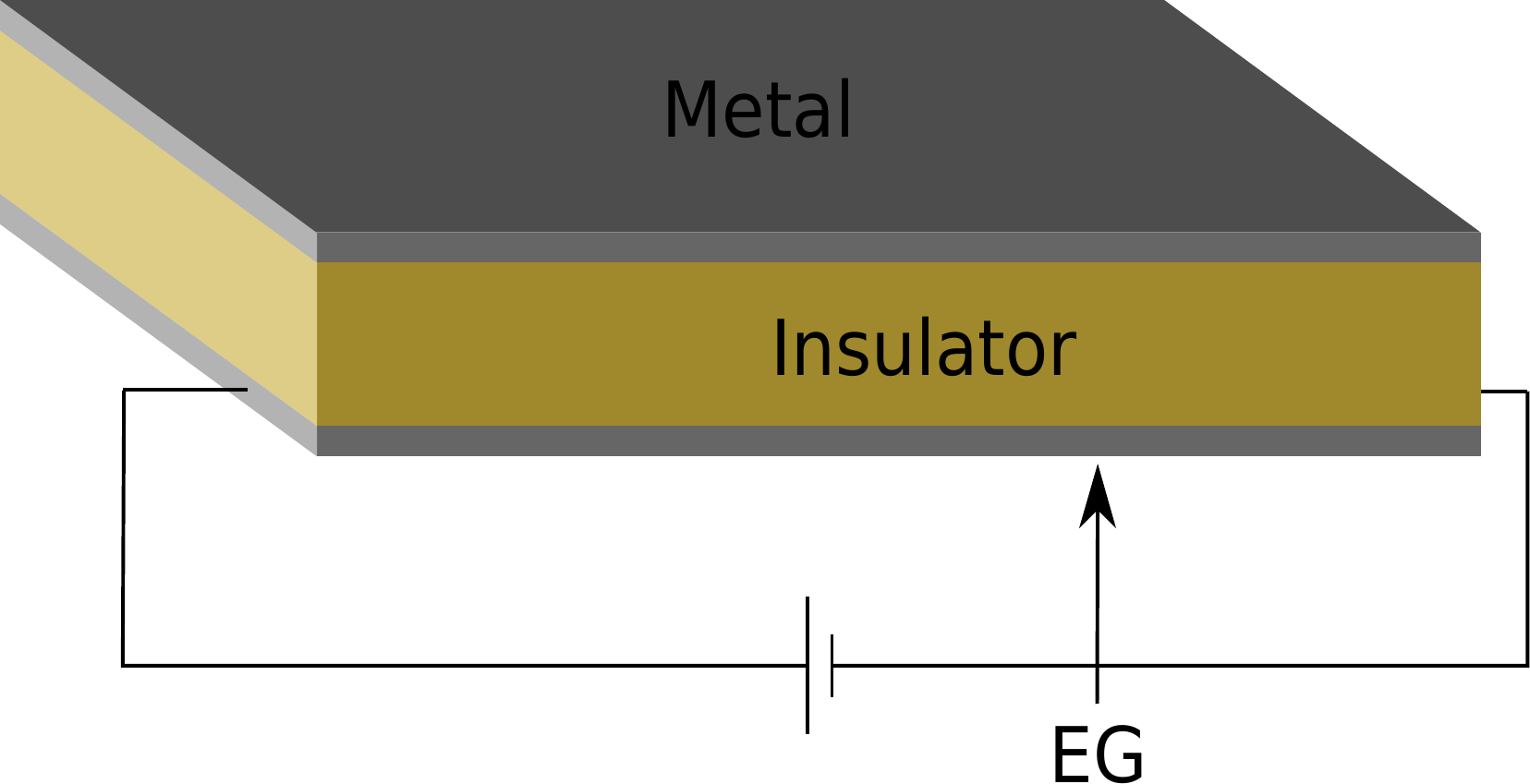}
\caption{\label{System} Illustration of the system.}
\end{figure}

We consider a 2D EG layer coupled to phonons and to electrons in a metal layer separated by an insulator of width $d$, as illustrated in Fig.~\ref{System}.
The Hamiltonian of the system reads
\begin{equation}
\label{ModelHamiltonian}
\begin{split}
H = &\sum_i \epsilon_i n_i + \frac{1}{2}\sum_{i\neq j} u_{ij}n_i n_j + \frac{1}{2}\sum_{i\neq j} c^{\dag}_i t_{ij} c_j\\
+ &\sum_q \omega_q a^{\dag}_q a_q + \sum_{i,q}g_{iq}\left(a^{\dag}_q + a_{-q}\right)n_i\\
+ &\sum_k E_k f^{\dag}_k f_k + \frac{1}{2}\sum_q V^{(2)}_q \rho_q \rho_{-q} + \sum_{i,q,\Omega} V^{(1)}_{q i} \rho_q n_i.
\end{split}
\end{equation}
Here $c_i$ ($c^{\dag}_i$) are the operators annihilating (creating) an electron in the EG at the localized
site $i$, $n_i \equiv c^{\dag}_i c_i$. The on-site energies $\epsilon_i$ are randomly distributed within the interval $[-W,W]$, and $t_{ij} \propto e^{-r_{ij}/\xi}$, typically small compared to $W$, represents the tunneling of electrons between the localized sites $i$ and $j$. Here $\xi$ is the localization length~\cite{MillerAbrahams1960}. The Coulomb interaction between sites $i$ and $j$
is given by $u_{ij}=\frac{e^2}{\kappa r_{ij}}$, where $\kappa$ is the dielectric constant. The operators $f_k$ and $f^{\dag}_k$ stand for the conduction 
electrons in the metallic plate and $E_k$ is the free electron energy with wave number $k$. The electron density in the metallic plate is given by $\rho_{q} =  \sum_{k} f^{\dag}_k f_{k-q}$. 
The bare Coulomb interaction in the metal is given by $V^{(2)}_q = \frac{2\pi e^2}{L^2 q}$. The Coulomb coupling between the EG and the metallic plate  is described by: 
\begin{equation}
\label{EG-Metal}
V^{(1)}_{qi} = \frac{2\pi e^2}{\kappa L^2 q}e^{-i\bm{qr}_i-qd} = V^{(1)}_q e^{-i\bm{qr}_i}\, ,
\end{equation}
where $\bm{r}_i$ is the location of site $i$ in the two dimensional EG, and $d$ is the distance between the metal layer and the EG system which we denote also as the layer separation. Finally, 
$a_q$ and $a^{\dag}_q$ describe phonons and $g_{qi} = g_q e^{-i\bm{qr}_i}$ is the electron-phonon (el-ph) interaction in the deformation potential approximation, $|g_q| \propto \sqrt{|q|}$.

Starting from the microscopic model we wish to derive an effective action for the EG degrees of 
freedom. We consider the partition function ${\cal Z} = \int D\bar\Psi D\Psi \exp[-\mathcal{S}]$, where 
$\Psi$ represents symbolically all the fermionic and bosonic fields in the problem.  
The action $\mathcal{S}$ can be obtained by performing the Legendre transform,\cite{altland_simons_2010,Tsvelik_2003},
\begin{equation}
\label{Legendre}
\mathcal{S}[\bar{\Psi},\Psi] = \int_0^{1} d\tau \left[ \bar{\Psi}\p_{\tau}\Psi + H(\bar{\Psi},\Psi)\right]\ .
\end{equation}
Here and throughout the paper $\hbar=1$, $k_B=1$, the energy (frequency) is measured in units of temperature and the imaginary time in units of 
inverse temperature. The proper units are reinstalled in the final results.

The microscopic action is composed of four parts:
\begin{equation}
\label{ModelAction}
\mathcal{S} = \mathcal{S}_\text{EG} + \mathcal{S}_t + \mathcal{S}_\text{ph}+\mathcal{S}_\text{el}.
\end{equation}
Here $\mathcal{S}_\text{EG}$ describes the on-site energies and the Coulomb interaction in the EG, 
whereas $\mathcal{S}_t$ describes tunneling in the EG: 
\begin{equation}
\begin{split}
\label{EG+Tunnelling}
\mathcal{S}_\text{EG} &= \sum_{i,\omega} \bar{c}_{i,\omega} (-i\omega + \epsilon_i) c_{i,\omega} + \frac{1}{2}\sum_{i\neq j, \Omega}u_{ij}\bar{n}_{i,\Omega}n_{j,\Omega}\, ,\\
\mathcal{S}_t &= \frac{1}{2}\sum_{i\neq j, \omega}\bar{c}_{i,\omega} t_{ij} c_{j,\omega}\, .
\end{split}
\end{equation}
The Matsubara Fourier transforms are defined as 
$c_{i,\omega} = \int_0^{1} d\tau \, e^{i\omega \tau} c_{i}(\tau)$ and
$n_{i,\Omega} = \int_0^{1} d\tau \, e^{i\Omega \tau} n_{i}(\tau) = \sum_{\omega}\bar{c}_{i,\omega}c_{i,\omega-\Omega}$, where $\omega$ denotes the fermionic Matsubara frequencies $(2\pi +1)n$
and $\Omega$ are the bosonic ones $\Omega=2\pi m$.

The phonons and their coupling to the EG are described by:
\begin{equation}
\begin{split}
\label{Phonons+Coupling}
\mathcal{S}_\text{ph} &=\sum_{q,\Omega} \bar{a}_{q,\Omega} (-i\Omega + \omega_q) a_{q,\Omega} \\
&\quad + \sum_{i,q,\Omega}g_{iq}\left(\bar{a}_{q,-\Omega} + a_{-q,\Omega}\right)n_{i,\Omega}\, .
\end{split}
\end{equation}
Since phonons can propagate also through the insulator and the substrate we consider a three dimensional phonon DOS. The two dimensional metal layer is represented by the Jellium model: 
\begin{equation}
\begin{split}
\label{Metal+Coupling+Int}
\mathcal{S}_\text{el} &= \sum_{k,\omega} \bar{f}_{k,\omega}(-i\omega+E_k)f_{k,\omega} + \frac{1}{2}\sum_{q,\Omega} V^{(2)}_q \bar{\rho}_{q,\Omega} \rho_{q,\Omega}\\
&\quad +\sum_{i,q,\Omega} V^{(1)}_{q i}\bar{\rho}_{q,\Omega} n_{i,\Omega}\, .
\end{split}
\end{equation}

The presence of the density-density interaction in the metal allows us to systematically derive the screening of the EG-Metal interaction ($V^{(1)}$) which results from the response of the metal electrons to the localized electrons in the EG system.

In what follows we derive the effective action and conductivity for a ballistic metal layer. In Sec.~\ref{WeakDisorder} we show how disorder in the metal is an important addition that can cause a substantial effect on the conductivity.

\subsection{Microscopic description of the electromagnetic and phononic fluctuations}

The first step in calculating the effective action is to eliminate the Coulomb interactions in the metal via the Hubbard-Stratonovich transformation and then integrate over the metal's electron fields \cite{altland_simons_2010, fradkin_2013}: 
\begin{equation}
\label{S_OmegaHubbard+Trln}
\begin{aligned}
\mathcal{S} &= \mathcal{S}_\text{EG} + \mathcal{S}_t +\mathcal{S}_\text{ph} + \frac{1}{2} \sum_{q,\Omega} \bar{\Phi}_{q,\Omega} \left(V^{(2)}_q\right)^{-1} \Phi_{q,\Omega}\\
&\quad - \text{Tr\,ln}\left(-G_{V^{(1)},\Phi}^{-1}\right)\, ,
\end{aligned}
\end{equation}
where:
\begin{equation}
\label{GeneralizedGreenFunc}
G_{V^{(1)},\Phi}^{-1} \equiv G^{-1}_0 - \bar{V}^{(1)}\bar{n} - i\bar{\Phi}.
\end{equation}
Here $G^{-1}_0$ is the inverse propagator of the free electron in the metal with matrix elements $ G^{-1}_{0k\omega} = i\omega - E_k$, $\bar{\Phi}_{k-k',\omega-\omega'} = \bar{\Phi}_{q,\Omega} = \Phi_{-q,-\Omega}$ is the potential field in the metal (introduced by the Hubbard-Stratonovich transformation) and the coupling to the EG system is represented by the matrix $\bar{V}^{(1)}n$ with the matrix elements $\bar{V}^{(1)}_{k-k' i} n_{i,\omega-\omega'} = \bar{V}^{(1)*}_{q i} \bar{n}_{i,\Omega} = \bar{V}^{(1)}_{-q i} n_{i,-\Omega}$.

Eq.~(\ref{S_OmegaHubbard+Trln}) is exact but not solvable. Here we consider the mean-field (MF) approximation for the field $\Phi$. In Appendix \ref{Fluctuations} we validate the MF approximation by considering fluctuations around the MF solution. The MF solution 
$\Phi^0$ solves the MF equation:
\begin{equation}
\label{Mean-Field}
0 = \frac{\delta \mathcal{S}}{\delta \bar{\Phi}^0_{q,\Omega}} = \left(V^{(2)}_q\right)^{-1} \Phi^0_{q, \Omega} + 2i\sum_{k,\omega} \left(G_{V,\Phi^0}\right)_{(k,\omega),(k + q, \omega + \Omega)}.
\end{equation}
Due to the presence of the EG contributions in (\ref{GeneralizedGreenFunc}) this is still a complicated equation to solve. We assume the metal to be an almost perfect screener within itself. This means the 
total potential in (\ref{GeneralizedGreenFunc}) must be small. We denote this total potential 
$\Delta\Phi^0$, i.e., 
\begin{equation}
\label{ansatz2}
i\Delta\Phi^0_{q,\Omega} = i\Phi^0_{q,\Omega} + \sum_i V^{(1)}_{qi}n_{i,\Omega}\, ,
\end{equation}
and expand the propagator (\ref{GeneralizedGreenFunc}) in $\Delta\Phi^0$.
Expanding Eq.~(\ref{GeneralizedGreenFunc}) to the linear order in $\Delta\Phi^0$
and substituting this to the MF equation (\ref{Mean-Field}) we obtain the following
MF solution (for details see Appendix~\ref{MFE-Sol}):
\begin{equation}
\label{MFE-ApproxSol}
i\Phi^0_{q,\Omega}  = -\sum_i (1 - f_{q\Omega})V^{(1)}_{qi}n_{i,\Omega}\, .
\end{equation}
The function $f_{q\Omega}$ is found to be the inverse RPA dielectric function:
\begin{equation}
\label{Deviation}
f_{q\Omega} = 1/\epsilon^{RPA}_{q\Omega} = \frac{1}{1-V^{(2)}_{q\Omega}\Pi_{q\Omega}}\, ,
\end{equation}
with the polarization function,
\begin{equation}
\label{Polarization}
\Pi_{q\Omega} = 2\sum_{k, \omega} G_{0k\omega}G_{0k+q,\omega + \Omega} = 2\sum_k \frac{N_{k,k+q}}{E_{k,k+q} + i\Omega}.
\end{equation}
Here $N_{k,k+q} \equiv N_{k}-N_{k+q}$, where $N_k$ is the Fermi occupation of state with energy $E_k$ in the metal, and $E_{k,k+q} = E_{k}-E_{k+q}$. 
Note that the MF potential $\Phi^0$ is a dynamical one due to the dynamics of the localized charges 
$n_i(\tau)$.

We can now check how justified was the expansion to the linear order in $\Delta\Phi^0$.
From Eq.~(\ref{MFE-ApproxSol}) we obtain
\begin{equation}
\label{DeviationField}
\Delta\Phi^0_{q,\Omega} = \sum_i f_{q\Omega}V^{(1)}_{qi}n_{i,\Omega}.
\end{equation}
In the static long wavelength limit we have
\begin{equation}
\label{ProofSmallDeltaPhi}
\begin{split}
&f_{q\Omega} \approx \frac{q}{q+q_{TF}} \approx \frac{q}{q_{TF}} < \frac{1}{q_{TF}d} \ll 1\, ,
\end{split}
\end{equation}  
with $q_{TF}=2/a_B$, $a_B$ is the Bohr radius and the long wavelength expansion is defined as:
\begin{equation}
\label{LongWaveLengthLimit}
E_{k,k+q} \approx \bm{k}\cdot \bm{q}/m; \; N_{k,k+q} \approx -\delta\left(k^2-k_F^2\right) 2 \bm{k}\cdot \bm{q}\, ,
\end{equation}
where $k_F$ is the Fermi wavenumber. Evidently from Eq.~(\ref{EG-Metal}), the EG-Metal separation serves as a cutoff for the e-h wavelengths $qd<1$ which justifies the long wavelength approximation. 
Thus the expansion is justified provided that the layer separation is large enough such that maximum relevant wavenumber ($1/d$) is much smaller than Thomas-Fermi wave number, $q_{TF}d \gg 1$ (an inequality which we consider throughout).

Substituting now $\Phi^0$ into Eq.~(\ref{S_OmegaHubbard+Trln}) and expanding $
\text{Tr\,ln}$ up to 
the second order in $\Delta \Phi_0$ we obtain the following MF action 
\begin{equation}
\begin{split}
\label{S_MF2}
&\mathcal{S}_\text{MF} = \mathcal{S}_\text{EG} + \mathcal{S}_t - \text{Tr\,ln}\left(-G^{-1}_0\right) \\
&\quad - \frac{1}{2} \sum_{i,j,\Omega,q} (1-f_{q\Omega})^2\left(V^{(2)}_q\right)^{-1} V^{(1)*}_{qi}V^{(1)}_{qj}\bar{n}_{i,\Omega}n_{j,\Omega}\\
&\quad + \frac{1}{2}\sum_{i,j,\Omega,q}\Pi_{q\Omega}f^2_{q\Omega}V^{(1)*}_{qi}V^{(1)}_{qj}\bar{n}_{i,\Omega}n_{j,\Omega}\, ,
\end{split}
\end{equation}
which gives [using (\ref{Deviation})]
\begin{equation}
\begin{split}
\label{S_MF3}
&\mathcal{S}_\text{MF} = \mathcal{S}_\text{EG} + \mathcal{S}_t + \frac{1}{2} \sum_{ij\Omega} K^{el}_{ij\Omega}\bar{n}_{i,\Omega}n_{j,\Omega}\, ,
\end{split}
\end{equation}
with
\begin{equation}
\label{ScreenedKernel}
K^{el}_{ij\Omega} = \sum_q f_{q\Omega}\Pi_{q\Omega}V^{(1)*}_{qi}V^{(1)}_{qj}\ .
\end{equation}
The constant $\text{Tr\,ln}\left(-G^{-1}_0\right)$ has been dropped.

For a non-interacting metal one would obtain instead of (\ref{ScreenedKernel}) a kernel of the 
form, $\sum_q \Pi_{q\Omega}V^{(1)*}_{qi}V^{(1)}_{qj}$. Therefore we denote $K$ in Eq.~(\ref{ScreenedKernel}) as the screened kernel. As expected, for $d=0$ the screened kernel together with the EG interaction give the known RPA interaction between impurities in the metal, $u_{ij}+K^{el}_{ij\Omega}=\sum_{q}f_{q\Omega}V^{(2)}_q e^{-i\bm{qr}_{ij}}$ which further validates our MF approximation. Given the typical distance between sites in the EG system being much larger than the Thomas-Fermi wavelength, the RPA is a good approximation. 

The screened kernel (\ref{ScreenedKernel}) can be alternatively derived by expanding Eq.~(\ref{S_OmegaHubbard+Trln}) to 2nd order in $V^{(1)}+i\Phi$ and performing a Gaussian integration over $\Phi$; this means that fluctuations of the field $\Phi$ are taken into consideration already at the MF level. Nevertheless, the MF analysis has a faster convergence than the perturbative loop expansion, at least for the given model, thus the MF approach should be useful for the calculation of higher order corrections.

Reinstalling back the phonon action, $\mathcal{S}_\text{ph}$, and integrating over the phonons degrees of freedom the total action reads:
\begin{equation}
\label{SeffPhiFinal2MediatedInt}
\mathcal{S}_\text{MF} = \mathcal{S}_\text{EG} + \mathcal{S}_t + \frac{1}{2}\sum_{i j \Omega}K_{ij\Omega}\bar{n}_{i , \Omega} n_{j,\Omega}\, ,
\end{equation}
with $K_{ij\Omega} = K^{el}_{ij\Omega} + K^{ph}_{ij\Omega}$, where $K^{ph}_{ij\Omega}$ is the phonon kernel:
\begin{equation}
\label{PhononKernel}
\begin{split}
&K^{ph}_{ij\Omega} = - \sum_q|g_q|^2 \frac{2\omega_q}{\omega_q^2 + \Omega^2}e^{i\bm{qr}_{ij}}\, ,
\end{split}
\end{equation}
and we use the identities $g_q = g^*_{-q}$ and $\omega_q = \omega_{-q}$. To see how the effective interaction renormalizes the EG parameters we split the kernel [Eq.~(\ref{ScreenedKernel}) and Eq.~(\ref{PhononKernel})] to a static and a dynamic part $K_{ij\Omega} = K_{ij0} + K'_{ij\Omega}$:
\begin{equation}
\begin{split}
\label{TotalKernel}
K_{ij0} &= \sum_q \left(V^{(1)}_q \right)^2 f_{q0} \Pi_{q0} e^{i\bm{qr}_{ij}} - 2\sum_q \frac{|g_q|^2}{\omega_q}e^{i\bm{qr}_{ij}}\\
K'_{ij\Omega} & = \sum_q \left(V^{(1)}_q\right)^2\left(f_{q \Omega}\Pi_{q,\Omega} - f_{q 0} \Pi_{q,0}\right)e^{i\bm{qr}_{ij}}\\
&\quad\; + 2\sum_q|g_q|^2 \frac{\Omega^2}{\omega_q(\omega_q^2 + \Omega^2)}e^{i\bm{qr}_{ij}}\\
&= \sum_q \left(V^{(1)}_q\right)^2 f_{q 0} f_{q \Omega}\left(\Pi_{q,\Omega} - \Pi_{q,0}\right)e^{i\bm{qr}_{ij}}\\
&\quad\; + 2\sum_q|g_q|^2 \frac{\Omega^2}{\omega_q(\omega_q^2 + \Omega^2)}e^{i\bm{qr}_{ij}}.
\end{split}
\end{equation}

Consequently the renormalized EG interaction (recasting back to units of energy) takes the known form:
\begin{equation}
\label{ScreendInt}
\tilde{u}_{ij} = u_{ij} + K_{ij0} \approx u_{ij} + K^{el}_{ij0} \approx \frac{e^2}{\kappa r_{ij}} - \frac{e^2}{\kappa \sqrt{r_{ij}^2+4d^2}}\, ,
\end{equation}
where $K^{ph}_{ij0}$ is neglected since we are interested in the case of weak el-ph interaction. Also, in the last step we used the long wavelength expansion Eq.~(\ref{LongWaveLengthLimit}) and the inequality $q_{TF}d \gg 1$. Eq.~(\ref{ScreendInt}) is the EG interaction screened by the presence of the metal~\cite{PhysRevB.49.13721,PhysRevB.51.16871}. As can be seen, the interaction $\tilde{u}_{ij}$ behaves as $1/r_{ij}$ for $r_{ij} \ll d$ and $1/r_{ij}^3$ for $r_{ij} \gg d$. 
The homogeneous shift of the on-site energies, due to $K_{ii0}$, is ignored.

\subsection{The dressed tunneling amplitude}

In this section we obtain the generic model of the dressed tunneling amplitude coupled to environmental modes (along the lines of Ref.\cite{NazarovIngold1992}). Performing the Hubbard-Stratonovich transformation to decouple the dynamical part of the interaction [last term in Eq.~(\ref{SeffPhiFinal2MediatedInt})] and a gauge transformation of the form,
\begin{equation}
c_i(\tau) \rightarrow e^{i\Theta_i(\tau)}c_i(\tau)\, ,
\end{equation}
we obtain our final effective action:
\begin{equation}
\label{ModelAction}
\begin{split}
\mathcal{S}_\text{eff} &= \tilde{\mathcal{S}}_\text{EG} + \tilde{\mathcal{S}}_t+\mathcal{S}_{\phi}\\
&=\tilde{\mathcal{S}}_\text{EG} + \int_0^{1} d\tau \sum_{i\neq j} \bar{c}_i(\tau) \tilde{t}_{ij} c_j(\tau) \\
&\quad + \frac{1}{2}\int_0^{1} \int_0^{1} d\tau d\tau' \sum_{i\neq j}\phi_i(\tau) \left(K'\right)^{-1}_{ij}(\tau-\tau') \phi_j(\tau')\, ,
\end{split}
\end{equation}
where $\tilde{\mathcal{S}}_\text{EG}$ is the EG action with the renormalized interaction [Eq.~(\ref{ScreendInt})], $\phi_i$ is the local potential on site $i$ introduced by the Hubbard-Stratonovich transformation, and the dressed tunneling amplitude is: 
\begin{equation}
\label{DressedTunneling}
\tilde{t}_{ij} = t_{ij} e^{i\Theta_{ij}(\tau)}\ ,
\end{equation} 
where $\Theta_{ij}(\tau) = \Theta_i(\tau)-\Theta_j(\tau)$ and $\Theta_i(\tau) = \int_0^{\tau} d\tau' \,\phi_i(\tau')$. 
The kernel $K'$ is given in Eq.~(\ref{TotalKernel}). Note that the potential field does not have a static part i.e. $\phi_{i,\Omega=0}=0$, since the kernel $K'_{ij,\Omega=0}=0$ by definition. Thus $\Theta_i(\tau)$ is periodic.

\section{Conductivity}
\label{ConductivitySec}

In this section we use our effective action [Eq.~(\ref{ModelAction})] to derive the DC conductance between sites $i$ and $j$ to leading order in the weak dressed tunnelling amplitude, $\tilde{t}_{ij}$. In Subsec.~\ref{WeakDisorder} we show how to generalize our results to the case of diffusive metallic plate. Finally, in Subsec.~\ref{MacroCond} we apply Mott's prescription to evaluate the dependence of macroscopic conductivity of the EG on temperature and use it to compare the conductivities with and without the metal layer.

We present here the main steps of the derivation of the conductance between sites $i$ and $j$, for further details see Appendix~\ref{GeneratingFunctionalApp}.
For the response function we obtain [see Eq.~(\ref{ConductivityKernel})]:
\begin{equation}
\begin{split}
\label{CondKernelExplicit}
&C^I_{ij}(\tau-\tau') = -\left\langle I_i(\tau)I_j(\tau')\right\rangle \Big|_{\chi=0} \\
&+ \delta(\tau-\tau')\sum_{l(\neq i)} \left(\delta_{ij} - \delta_{jl}\right)\left\langle\bar{c}_i(\tau) \tilde{t}_{i l}c_l(\tau) + \text{H.c.}\right\rangle\Big|_{\chi=0}\, ,
\end{split}
\end{equation} 
where
\begin{equation}
I_i(\tau) =\sum_{j(\neq i)} \left(\bar{c}_i(\tau) \tilde{t}_{i j}e^{i\chi_{i j\tau}}c_j(\tau) - \bar{c}_j(\tau) \tilde{t}_{i j}^{*}e^{-i\chi_{i j\tau}}c_i(\tau)\right)
\end{equation}
is the current entering (leaving) the site $i$ [defined in Eq.~(\ref{Current})] and $\chi_i(\tau) = \int_0^{\tau} d\tau' U_i(\tau')$ where $U_i(\tau)$ is the classical external potential field at site $i$ in the EG system. The first and second terms in Eq.~(\ref{CondKernelExplicit}) are called the paramagnetic and diamagnetic contributions, respectively. 

To find the conductance to order $|\tilde{t}|^2$ we evaluate the paramagnetic term to zeroth order in the dressed tunneling action, $\tilde{\mathcal{S}}_t$ (since the current operator is proportional to $\tilde{t}$) and the diamagnetic term we calculate to first order. Substituting Eq.~(\ref{DressedTunneling}) and calculating the averages explicitly we obtain:
\begin{equation}
\begin{split}
\label{FinalCondKernel}
C^I_{ij\Omega} &\approx \sum_l\left(\delta_{ij}-\delta_{jl}\right)|t_{i l}|^2\int_0^{1} d\tau (1-e^{i\Omega\tau})e^{J_{il}(\tau)} \times\\
&\quad\, \times\left(n_i(1-n_{l})e^{E_{il}\tau} + n_l(1-n_i)e^{-E_{i l}\tau}\right).
\end{split}
\end{equation}
Here $E_{il}=E_i-E_l$ where $E_i$ are the on-site energies, which take into account the EG Coulomb interactions screened by the metallic plate. These are already included in 
$\tilde{\mathcal{S}}_\text{EG}$ of Eq.~(\ref{ModelAction}). Furthermore, $J_{ij}$ is the correlation function of the gauge field $\Theta_{ij}=\Theta_i-\Theta_j$:
\begin{equation}
\label{J-Correlation}
J_{ij}(\tau) = \la\Theta_{ij}(\tau)\Theta_{ij}(0)\ra_{0\phi}-\la \Theta_{ij}(0)^2 \ra_{0\phi}\, ,
\end{equation}
where the average is done with respect to $\mathcal{S}_{\phi}$ defined in Eq.~(\ref{ModelAction}). Performing the analytical continuation to real time and frequency~\cite{Efetov2003} and then taking the limit $\omega \rightarrow 0$, we obtain the DC conductance between sites $i$ and $j$ (in dimensionfull units) given in Eq.~(\ref{Conductance}) with
\begin{equation}
\label{PofE}
P(E_{ij}) = \frac{1}{2\pi}\int_{-\infty}^{\infty} dt \; e^{J_{ij}(t) + iE_{ij}t}.
\end{equation}
From this point on - real times and energies/frequencies are in dimensionfull units.  
The correlation function of the Gauge field can be divided into the contributions of the electron and phonon environments respectively, $J_{ij}(t) = J^{el}_{ij}(t) + J^{ph}(t)$, which are given by (see derivation in Appendix~\ref{J-Evaluation}):
\begin{equation}
\begin{split}
\label{J-Realtime}
J_{ij}^{el}(t) &= \int_0^{\infty} \frac{d\omega}{\omega^2}S^{el}_{ij}(\omega) F(\omega,it)\, ,\\
J^{ph}(t) &= \int_0^{\infty} \frac{d\omega}{\omega^2}S^{ph}(\omega)F(\omega,it),
\end{split}
\end{equation}
with
\begin{equation}
\label{F-RealTime}
F(\omega,it) = \coth\left(\frac{\beta\omega}{2}\right)(\cos(\omega t)-1) - \sin(\omega t)\, ,
\end{equation}
where the long wavelength limit is performed. The spectral functions at low energies are given by:
\begin{equation}
\begin{split}
\label{Spectral}
S^{el}_{ij}(\omega) &\approx \alpha_{ij}\omega e^{-\omega/\omega_c},\\
S^{ph}(\omega) &\approx \frac{\omega^s}{\tilde{\omega}^{s-1}}\Theta(\omega^{ph}_{c} - \omega)\, ,
\end{split}
\end{equation}
where $s$ is the dimensionality of the phonon environment, $\omega_c$ and $\omega_c^{ph}$ are the cutoff frequencies of the EG-metal and el-ph interactions respectively and $\tilde{\omega}$ is an energy scale inversely proportional to the deformation potential. The dimensionless coupling constant to the electronic environment (metal layer) is:
\begin{equation}
\label{EHCoupling}
\alpha_{ij} = \frac{1}{4 \pi \kappa^2}\frac{1}{k_F d}\ln\left[\frac{1}{2}+\frac{1}{2}\sqrt{1+\left(\frac{r_{ij}}{2d}\right)^2}\right].
\end{equation}

The general form of Eq.~(\ref{Conductance}) is of course intimately related to the 
equilibrium transition rates $2\pi |t_{ij}|^2 n_i(1-n_j) P(E_{ij})$ obtained in~\cite{NazarovIngold1992}. Given that the bare tunneling amplitude is typically small compared to the disorder energy, the expression for the conductance given in Eq.~(\ref{Conductance}) is applicable for a wide range of coupling strengths of the phonon and metal environments. 
A Fourier transform of Eq.~(\ref{PofE}) gives $P(E_{ij})$ as a convolution of contributions of metallic plate and phonons and we obtain Eq.~(\ref{Convolution}). The conductance between two sites
with energies $E_I$ and $E_F$ separated by distance $r$ then reads 
\begin{eqnarray}
\label{Conductance2}
&&\sigma_{I\rightarrow F} = \nonumber \\  && 2\pi\beta |t(r)|^2 n_I(1-n_F)  \int_{-\infty}^{\infty} dE' P^{el}(E',r)P^{ph}(E-E')\ ,\nonumber\\
\end{eqnarray} 
where $n_I \equiv n_F(E_I)$ and $n_F\equiv n_F(E_F)$.
The electronic absorption/emission probability density reads:
\begin{equation}
\label{PofE-EH}
P^{el}(E) = \frac{1}{2\pi}\int_{-\infty}^{\infty} dt \; e^{J^{el}(t) + i E t}\, ,
\end{equation}
and similarly for the phononic one, $P^{ph}(E-E')$, with $J^{el}$ replaced by $J^{ph}$.
The phonon correlation functions can be further divided into two parts [see Eq.~(\ref{J-Correlation})], the Debye-Waller term and the rest,  $J^{ph} = -W^{ph} + \tilde{J}^{ph}$, where 
$W^{ph} \equiv \la \Theta^{ph}_{ij}(0)^2 \ra$.
For the electronic Ohmic environment such a division 
does not make sense as both parts would strongly diverge, whereas $W^{ph}$ has a finite value.
We thus choose to separate it from the phonon correlation function $J^{ph}$:
\begin{equation}
\label{PhononPofEAndDW}
P^{ph}(\Delta E) = e^{-W^{ph}}\frac{1}{2\pi}\int_{-\infty}^{\infty} dt \; e^{\tilde{J}^{ph}(t) + i \Delta E t},
\end{equation}
with $\Delta E = E - E'$ and $e^{-W^{ph}}$ is the well known Debye-Waller factor. The Debye-Waller exponent is given by:
\begin{equation}
\label{DWexponent}
W^{ph} = \frac{1}{\tilde{\omega}^{s-1}}\int_0^{\omega^{ph}_c} d\omega \; \omega^{s-2} \coth\left(\frac{\beta \omega}{2}\right)\, ,
\end{equation}
where we substitute the phonon spectral function given in Eq.~(\ref{Spectral}).
To obtain the single phonon assisted tunneling we expand Eq.~(\ref{PhononPofEAndDW}) to leading order in $\tilde{J}^{ph}$ and obtain:
\begin{equation}
\label{P-SinglePhonon}
\begin{split}
&P^{ph}(\Delta E) \approx e^{-W^{ph}}\int_{-\infty}^{\infty} \frac{dt}{2\pi} \; e^{i \Delta E t} (1 + \tilde{J}^{\varphi}(t)) \\
&= e^{-W^{ph}}\left\{\delta(\Delta E) + \frac{1}{\tilde{\omega}^{2}}|\Delta E|[n^{ph}(|\Delta E|)+\Theta(\Delta E)]\right\}\, ,
\end{split}
\end{equation}
where $n^{ph}$ is the phonon occupation in the EG system. As explained in the previous section we consider a 3D phonon spectral function [$s=3$ in Eq.~(\ref{Spectral})]. In that case we have $\tilde{\omega}^2 = \frac{2\pi \hbar^3 \rho c^5}{3\gamma^2}$ where $c$ is speed of sound, $\rho$ is the mass density and $\gamma$ is the deformation potential. Substituting Eq.~(\ref{P-SinglePhonon}) back in Eq.~(\ref{Conductance2}) we obtain our final form of the single-phonon conductance:
\begin{equation}
\label{CondSinglePh}
\begin{split}
&\sigma_{I\rightarrow F} = 2\pi\beta |t(r)|^2 n_I(1-n_F)e^{-W^{ph}}\left\{P^{el}(E,r)+\right.\\
&\left.\; + \frac{1}{\tilde{\omega}^2}\int_{-\infty}^{\infty}dE' \; P^{el}(E',r) |\Delta E| [n^{ph}(|\Delta E|)+\Theta(\Delta E)]\right\}\, ,
\end{split}
\end{equation}
where the arguments $r$, $E_I$ and $E_F$ are suppressed in Eq.~(\ref{CondSinglePh}) for compactness.
As can be seen, negative energy difference, $E < 0$, describes assisted tunneling while $E>0$ describes the dissipation to the e-h and phonon environments.
As expected, without the metal layer (setting $J^{el}=0$) the conductance reduces to the typical form given in usual analysis of resistor network~\cite{MillerAbrahams1960, Shklovskii1984}:
\begin{eqnarray}
\label{MilleAbrahams}
&&\sigma_{0,I\rightarrow F} = \nonumber \\ &&2\pi\beta |t(r)|^2 n_I(1-n_F)\frac{e^{-W^{ph}}}{\tilde{\omega}^2} |E| [n^{ph}(|E|)+\Theta(E)]\ .\nonumber\\
\end{eqnarray}
The resonant tunneling term does not contribute in Anderson insulators and therefore neglected in the calculation of $\sigma_0$. The phonon Debye-Waller factor is usually discarded in the resistor network analysis assuming it is of order unity. This is based on the assumption that the el-ph interaction is weak w.r.t disorder energy and the near neighbour Coulomb ineraction in the EG system. Additional reason is that the Debye-Waller exponent usually has an IR cutoff which further decreases $W^{ph}$. The IR cutoff can be estimated self consistently via variational calculation of the free energy~\cite{VarPolaron1,VarPolaron2,VarPolaron3}. Regardless, in what follows we calculate the ratio of the conductivities given in Eq.~(\ref{CondSinglePh}) and Eq.~(\ref{MilleAbrahams}), which is not dependent on the phonon Debye-Waller factor.

\subsection{Weak disorder in the metal}
\label{WeakDisorder}

In most realistic systems the metallic layer is diffusive. Thus, it would be interesting to estimate the e-h spectral function [Eq.~(\ref{Spectral})] and its dependence on the hopping distance in the presence of disorder. For weak disorder, $k_Fl \gg 1$, where $l$ is the mean free path in the metal, the polarization function can be estimated by the diffusion approximation~\cite{RevModPhys.57.287,RevModPhys.66.261}:  
\begin{equation}
\label{polarizationDisorder}
\Pi^D_{q\Omega} = -\frac{\nu L^2 Dq^2}{|\Omega|+Dq^2}\, ,
\end{equation}
where $\nu = m/\pi \hbar^2$ is the DOS of the electrons in the metal, $L$ is the size of the sample, and $D = v_F l/2$ is the diffusion constant. Eq.~(\ref{polarizationDisorder}) is applicable for large length and time scales,
\begin{equation}
\label{DiffusionApprox}
ql \ll 1, \; \Omega \tau_s \ll 1\, ,  
\end{equation}
where the scatting time is $\tau_s = l/v_F$ given the Fermi velocity $v_F$. The diffusive kernel for the disordered metal is then obtained by replacing $\Pi \rightarrow \Pi^D$ in the screened kernel given in Eq.~(\ref{ScreenedKernel}). The dynamical part of the kernel takes the form:
\begin{equation}
\label{DressedDisorderedKernnel}
\begin{split}
&{K'}^D_{ij\Omega} = \beta\sum_q \frac{\nu L^2 |\Omega|q}{(|\Omega|+Dqq_{TF})q_{TF}} V^{(1)*}_{qi}V^{(1)}_{qj}\, ,
\end{split}
\end{equation}
and its static part remains unchanged and therefore is given in Eq.~(\ref{TotalKernel}). To obtain the conductivity of the EG system in the presence of disorder in the metal we consider the correlation function $J^{el}$ of the electronic bath [see Eq.~(\ref{J-Realtime})] in three regimes: (1) $l > d$, (2) $0 < l < \sqrt{\frac{d}{q_{TF}}}$, and (3) $\sqrt{\frac{d}{q_{TF}}} < l < d$. Regime No.1 is dominated by a ballistic motion in the metal and therefore $J^{el}$ is given by Eq.~(\ref{J-Realtime}). Regime No.2 is dominated by the diffusive motion in the metal. For time and length scales obeying Eq.~(\ref{DiffusionApprox}), the conductivity is obtained by replacing the correlation function $J^{el}$ in Eq.~(\ref{PofE-EH}) by the disordered one, $J^D_{ij}(\tau) = \la\Theta^D_{ij}(\tau)\Theta^D_{ij}(0)\ra_{0\phi}-\la \Theta^D_{ij}(0)^2 \ra_{0\phi}$. 

To calculate $J^D$ we use the diffusive kernel [Eq.~(\ref{DressedDisorderedKernnel})] instead of the screed kernel in the action [Eq.~(\ref{ModelAction})] and then repeat the steps done in Appendix~\ref{J-Evaluation}.
As can be seen, the diffusive kernel is not analytic, thus to perform the Matsubara summation we choose a contour of integration that avoids the real frequency axis. For the upper complex plane the integration contour is a semicircle shifted slightly above the real axis with radius $R\rightarrow \infty$. Similarly for the lower plane the contour is slightly shifted below the real axis. Following these steps we find that the correlation function $J^D$ has the same form as in Eq.~(\ref{J-Realtime}), i.e.:
\begin{equation}
\label{JD-Correlation}
J^D_{ij}(t) = \int_0^{\infty} \frac{d\omega}{\omega^2}S^D_{ij}(\omega) F(\omega,t)\, ,
\end{equation}
with the spectral function,
\begin{equation}
\label{SpectralFuncDisorderInt}
S^D_{ij}(\omega) = \frac{\omega}{k_F l \kappa^2} G^D_{ij}(\tilde{x}_1)\, ,
\end{equation}
and a form factor,
\begin{equation}
\label{GfuncDisInt}
G^D_{ij}(\tilde{x}_1) = \int_0^{\infty}dx\frac{x e^{-2x}}{x^2+\tilde{x}_1^2}[1-J_0(x R_{ij})].
\end{equation}
Here $x = qd$, $\tilde{x}_1 = \omega/\omega_1$, $\omega_1 = Dq_{TF}/d = v_F\frac{q_{TF}l}{2d}$ is the cutoff frequency of the EG-metal interaction, $R_{ij}=r_{ij}/d$, $J_0$ is the Bessel function and the inequality $(q_{TF}d) \gg 1$ is used. 
The low energy behaviour of the spectral function is Ohmic with the dimensionless coupling constant:
\begin{equation}
\label{DEHCoupling}
\alpha^D_{ij} = \frac{G^D_{ij}(0)}{k_F l \kappa^2} = \frac{1}{k_F l \kappa^2}\ln\left[\frac{1}{2}+\frac{1}{2}\sqrt{1+\left(\frac{r_{ij}}{2d}\right)^2}\right].
\end{equation}
For simplicity, in the next section we model Eq.~(\ref{GfuncDisInt}) with an exponential cutoff [see discussion blow Eq.~(\ref{PofE-EH-Ohmic})].
Comparing Eq.~(\ref{DEHCoupling}) and Eq.~(\ref{EHCoupling}) one can see that the coupling with and without the disorder has the same dependence on the hopping distance, $r_{ij}$ at the low frequency limit. 

Finally, in the intermediate regime No.3, the diffusive and ballistic contributions are comparable.  
The qualitative behaviour of the conductivity is obtained by using the diffusive correlation function [Eq.~(\ref{JD-Correlation})] for time and length scales obeying Eq.~(\ref{DiffusionApprox}) and the ballistic correlation function [Eq.~(\ref{J-Realtime})] for short time and length scales:
\begin{equation}
\begin{split}
\label{JBal+Diff}
J^{el}_{ij}(t) &= \frac{1}{k_F l \kappa^2}\int_0^{\frac{2d}{l^2 q_{TF}}} \frac{d \tilde{x}_1}{\tilde{x}_1}G^D_{ij}(\tilde{x}_1) F(\tilde{x}_1,t)\\
& \quad + \frac{1}{4\pi k_F d \kappa^2}\int_{\frac{d}{l}}^{\infty} \frac{d x_1}{x_1} G_{ij}(x_1) F(x_1,t).
\end{split}
\end{equation}
Here $1/l$ serves as the upper cutoff of the diffusive form factor, $G^D_{ij}(\tilde{x}_1) = \int_0^{\frac{d}{l}} dx \frac{x e^{-2x}}{x^2 + (\tilde{x}_1)^2}(1-J_0(x R_{ij}))$, and the lower cutoff for the ballistic form factor, $G_{ij}(x_1) = \int_{\frac{d}{l}}^{\infty}dx \frac{e^{-2\sqrt{x^2 + x_1^2}}}{\sqrt{x^2 + x_1^2}}(1-J_0(\sqrt{x^2 + x_1^2} R_{ij}))$. We assume that the crossover between the ballistic and diffusive regimes are captured qualitatively by the itegration limits~\cite{kamenev_2011} (note that $\tilde{x}_1/x_1 = q_{TF}l$ and therefore the integration limits of the ballistic and diffusive contributions are complementary).

Comparing the dimensionless Ohmic coupling constants given in Eq.~(\ref{JBal+Diff}) one can show that even in regime No.3 (i.e. $\sqrt{\frac{d}{q_{TF}}} < l < d$) the diffusive contribution dominates and the ballistic contribution can be safely neglected. Bearing this in mind we continue to the next section considering specifically the diffusive case ($l/d \lesssim 1$) for the calculation of the macroscopic VRH conductivity.

An accurate description for the ballistic-diffusive crossover (given in Ref.~\cite{BallisticDiffusiveCrossover2001}) coincides with our result in the diffusive limit, which is the relevant regime in our work, as stated above.

\subsection{Qualitative estimation of the polaronic reduction of the conductivity}
\label{MacroCond}

To estimate the conductivity we consider the diffusive regime of the metallic plate where the mean free path is smaller than the EG-Metal separation, $l/d \lesssim 1$. Using Mott's method~\cite{N.F.Mott} we calculate the ratio $\sigma(T)/\sigma_0(T)$ as a function of temperature, where $\sigma$ and $\sigma_0$ are respectively the conductivities with and without the presence of the metal layer. 

We start from evaluating Eq.~(\ref{PofE-EH}). Using Eq.~(\ref{Spectral}) and invoking the scaling limit, $\beta \omega^D_c, \omega^D_c t \gg 1$, we get~\cite{Leggett1987,kagan1987quantum}:
\begin{equation}
\label{PofE-EH-Ohmic}
P^{el}(E) \approx \frac{1}{2\pi\omega^D_c}\frac{e^{E/2T}}{\Gamma(\alpha_r)}\left|\Gamma\left(\frac{\alpha_r}{2}+\frac{iE}{2\pi T}\right)\right|^2 \left(\frac{\omega^D_c}{2\pi T}\right)^{1-\alpha_r}\, ,
\end{equation}
where $E$ is the energy difference between the initial and final states, $\alpha_r$ is the EG-Metal dimensionless coupling strength given in Eq.~(\ref{DEHCoupling}), $r_{ij} = r$, and $\Gamma$ is the Gamma function. Since we are interested in the low energy behaviour we approximate for simplicity an exponential cutoff to the diffusive spectral function [Eq.~(\ref{SpectralFuncDisorderInt})], $e^{-\omega/\omega^D_c}$. The upper frequency cutoff can be roughly estimated to be:
\begin{equation}
\label{DiffusionCutoffFerquency}
\omega^D_c \approx \text{min}\left(\omega_1,\;2\pi/\tau_s,\; 2\pi/\tau_{BL}\right)\sim 10^3 K.
\end{equation}
The three different frequencies are: (1) $\omega_1$ is the cutoff frequency of the EG-Metal interaction [see Eq.~(\ref{GfuncDisInt})], (2) $2\pi/\tau_s$ is the diffusion cutoff. Environmental modes with higher frequecies have a ballistic motion, and (3) $2\pi/\tau_{BL}$ is the tunneling frequency, where $\tau_{BL} = r/v_B$ is the B\"uttiker-Landauer tunneling time~\cite{ButtikerLandauer}, $r$ is the hopping length and $v$ is the imaginary velocity determined in the inverted potential barrier. Environmental modes with higher frequencies than the tunnelling frequency respond adiabatically to the tunneling electron and should not be included in the polaronic response. Note that one can estimate the typical $v$ to be associated with tunneling barriers of the order of the disorder energy $W$~\cite{DisorderStrengthZvi}.
It is argued in Refs.~\cite{WlargerThenFermi1,WlargerThenFermi2,DisorderStrengthZvi} that the typical disorder energy is larger than the Fermi energy in the EG system, thus $v$ is generally larger than the Fermi velocity.

\begin{figure}
\includegraphics[scale=0.28]{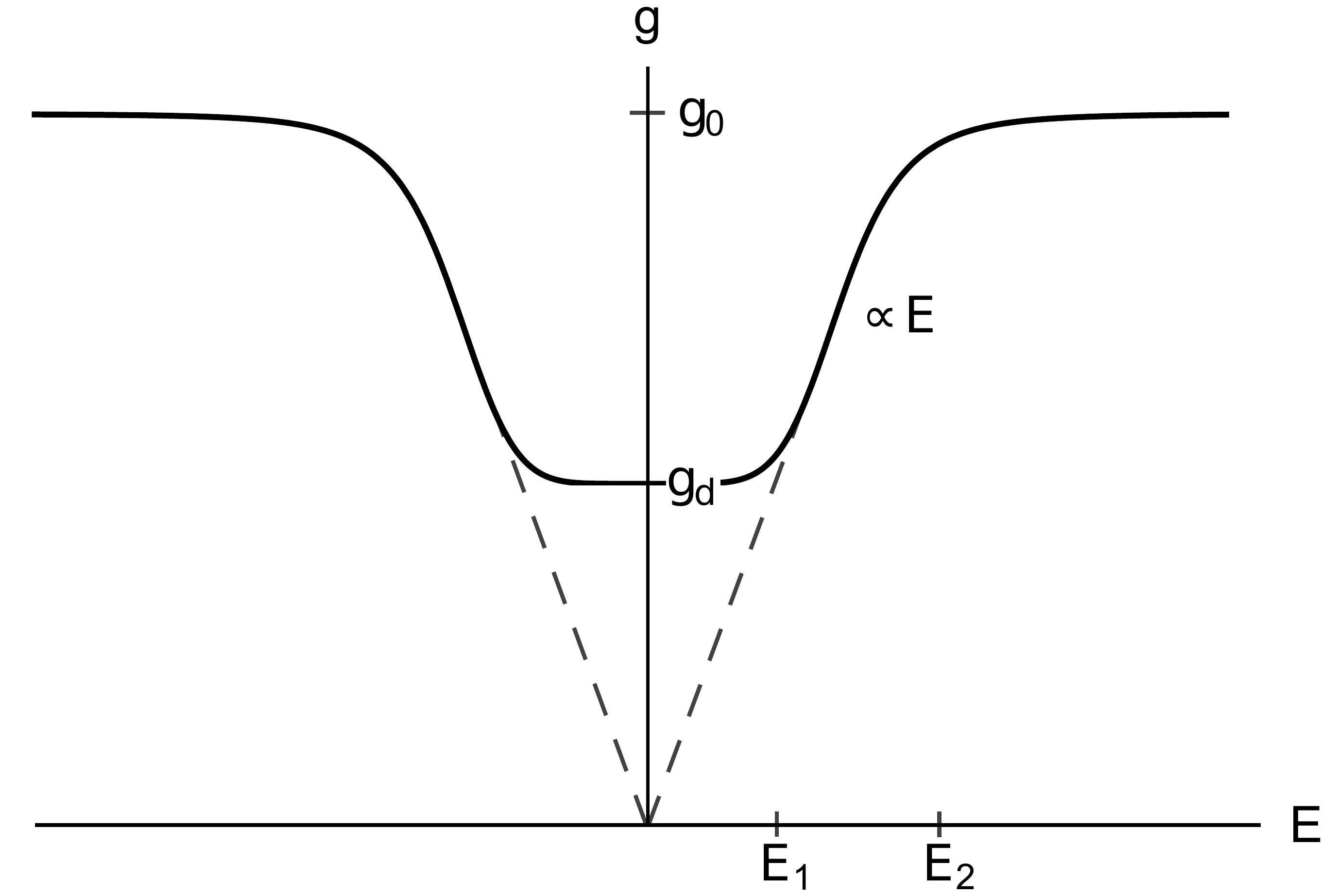}
\caption{\label{DOS-Schematics} Schematic description of the DOS in the presence of a metal layer in units of $\frac{\kappa^2}{e^4}$. The three regimes are: (1) Constant DOS at high energies, $g_0$, for $E > E_2 \approx g_0\frac{(e^2/\kappa)^2}{2/\pi}$ \cite{Efros_1975}, (2) Constant DOS at low energies, $g_d = 0.1(\kappa/e^2 d)$~\cite{PhysRevB.49.13721} for $E<E_1 \approx (g_d/g_0)E_2$, and (3) ES DOS at intermediate values (i.e. linear with the energy). The dashed line indicates ES DOS without the metal layer.}
\end{figure}

The energies $E$ in Eq.~(\ref{PofE-EH-Ohmic}) are distributed according to the DOS $g(E)$ of the electrons in the EG, described by the Hamiltonian:
\begin{equation}
\label{tildeH_EG}
\tilde{H}_\text{EG} = \sum_i \epsilon_i n_i + \frac{1}{2}\sum_{i\neq j}\frac{e^2}{\kappa}\left(\frac{1}{r_{ij}} - \frac{1}{\sqrt{r_{ij}^2+4d^2}}\right)n_in_j\, ,
\end{equation}
where the interaction term is given in Eq.~(\ref{ScreendInt}). $\tilde{H}_\text{EG}$ is obtained from the Lagendre transform of $\tilde{\mathcal{S}}_\text{EG}$ given in Eq.~(\ref{ModelAction}).
For the screened Coulomb interaction given in Eq.~(\ref{tildeH_EG}) the DOS has three distinct regions as can be seen schematically in Fig.~\ref{DOS-Schematics}. 

The crossover enegy between the constant DOS, $g(E) = g_0$, at high energies to the gapped ES DOS, $g(E)\propto|E|$, is given by the width of the ES gap, i.e. $E_2\approx  g_0\frac{(e^2/\kappa)^2}{2/\pi}$\cite{Efros_1975}. 
The screening of the metal results in a second (lower) crossover energy, $E_1$, to constant DOS $g(0)\propto e^2/\kappa d$~\cite{PhysRevB.49.13721}, up to logarithmic accuracy. $E_1 \approx (g_d/g_0)E_2$ is obtained similarly to the approach used in Ref.~\cite{Efros_1975}.

To find the VRH exponent ($p$) in a given regime we use Mott's method~\cite{N.F.Mott,larkin1982activation,Shklovskii1984} which goes as follows: Assuming the temperature is sufficiently lower than the characteristic energy difference of two localised sites ($E \gg T$) one can approximate the conductance between two localized sites as an exponential, $\sigma \propto e^{-h(E,r)}$ where $r$ is the hopping distance and $E$ serves as the effective hopping energy difference. After representing $E$ in terms of $r$, one defines the optimal hopping distance $\tilde{r}(T)$ as the minimum point of $h(E(r),r)$. $\tilde{r}$ is then substituted back to the exponent which results in the known VRH form given in Eq.~(\ref{VRH-form}). Repeating these steps for the conductance in the presence of the polaronic effect [Eq.~(\ref{CondSinglePh})], we find numerically a small deviation from the $\tilde{r}$ obtained by ES's and Mott's VRH (for localization lengths not too large, $\xi \lesssim d$). Therefore one can conclude that polaronic effect has a small effect on the exponent $p$. This can be explained by the fact that the coupling $\alpha_r$ has logarithmic dependence on $r/d$ (for $r > d$) which is weaker than the $r$ dependence of $E(r)$, i.e. $E(r) \propto 1/r$ for ES DOS~\cite{larkin1982activation} and $E(r) \propto 1/r^2$ for constant DOS~\cite{N.F.Mott}. For $r < d$ the coupling, $\alpha(r)$, is sufficiently weak which also results in a small polaronic effect. Our numerical evaluation is conducted only for the assisted hopping process ($E < 0$), a process which serves as the bottleneck of the conductance. Since the dynamical polaronic effect does not affect appreciably the ES-Mott crossover we consider its effects in the ES and Mott VRH regimes separately. In the ES VRH we use the optimal hopping length given without taking into account the polaronic effect, $r_{ES}$\cite{Efros_1975,larkin1982activation,Shklovskii1984}:
\begin{equation}
\begin{split}
\label{Optimal-ES-r}
\tilde{r}(T) &= r_{ES}(T) = \left(\frac{\pi}{2}\right)^{1/4}\xi\sqrt{\frac{U_{\xi}}{2T}}\, ,\\
E_{ES}(T) &= E(r_{ES}(T)) = \left(\frac{\pi}{2}\right)^{1/4}\sqrt{2U_\xi T}\, ,
\end{split}
\end{equation}
where $U_{\xi} = e^2/\kappa \xi$ and we used the relation $E_{ES}(T)=\sqrt{\frac{\pi}{2}}\frac{e^2}{\kappa r_{ES}(T)}$.

Mott's and ES's VRH arises from constant and gapped DOS respectively. Thus the structure of the DOS as depicted above gives rise to a Mott-ES-Mott crossovers. We denote these as the low and high temperature crossovers.
The ES VRH is given in the temperature range $T_1< T <T_3$. The low temperature ES-Mott cross over is given by~\cite{PhysRevB.49.13721}:
\begin{equation}
\label{LT-Crossover}
T_1 = (U_d/75)(\xi/ d)\, ,
\end{equation}
with $U_d = e^2/\kappa d$. The high temperature ES-Mott crossover is given by~\cite{Efros_1975,AharonyES-MottCrossover},
\begin{equation}
\label{HT-Crossover}
T_3 = \frac{\sqrt{\pi^3}}{2}(g_0\xi^2)^2U_{\xi}^3.
\end{equation}
For temperatures lower than $T_1$, the constant DOS near Fermi energy is larger than without the metal layer. This contributes to the increase in the conductivity\cite{PhysRevB.49.13721} and counteracts the reduction caused by the polaronic effect. However we consider the intermediate, ES regime where the increase of the DOS as a result of the screening is negligible. In this regime the polaronic effect is dominant at temperatures lower than $T_2$ where the hopping length is comparable to $d$, i.e., $r_{ES}(T_2) = d$ which gives $T_2 = \frac{1}{2}\sqrt{\frac{\pi}{2}}U_d\frac{\xi}{d}$. For temperatures higher than $T_2$ we have $r_{ES}(T) < d$, in this range the coupling $\alpha_r$ is small and decreases as $r^2$ [see Eq.~(\ref{DEHCoupling})]. Furthermore, a dominant polaronic effect w.r.t the static effect is obtained for $T_2 > T_1$, which is in agreement with parameters of the experiment~\cite{ZviCommunication, PhysRevLett.56.643}. This range of temperatures is given by (for $T_2 < T_3$):
\begin{equation}
\label{DominancePolaron}
\frac{1}{75} < \frac{T}{U_d}\frac{d}{\xi} \lesssim \left(\frac{\pi}{2^3}\right)^{1/2},
\end{equation}
which is compatible with the experiment's entire temperature range~\cite{PhysRevLett.56.643,ZviCommunication} (for $\xi \lesssim d$). This is since the increase of the DOS is small in dielectrics, unlike in the case of disordered semiconductors where a total increase in the conductivity is observed~\cite{HU199565} for $T < T_1$. We therefore assume henceforth that within the regime given by Eq.~(\ref{DominancePolaron}) the static effect is small and therefore neglected.

\begin{figure}
\includegraphics[scale=0.335]{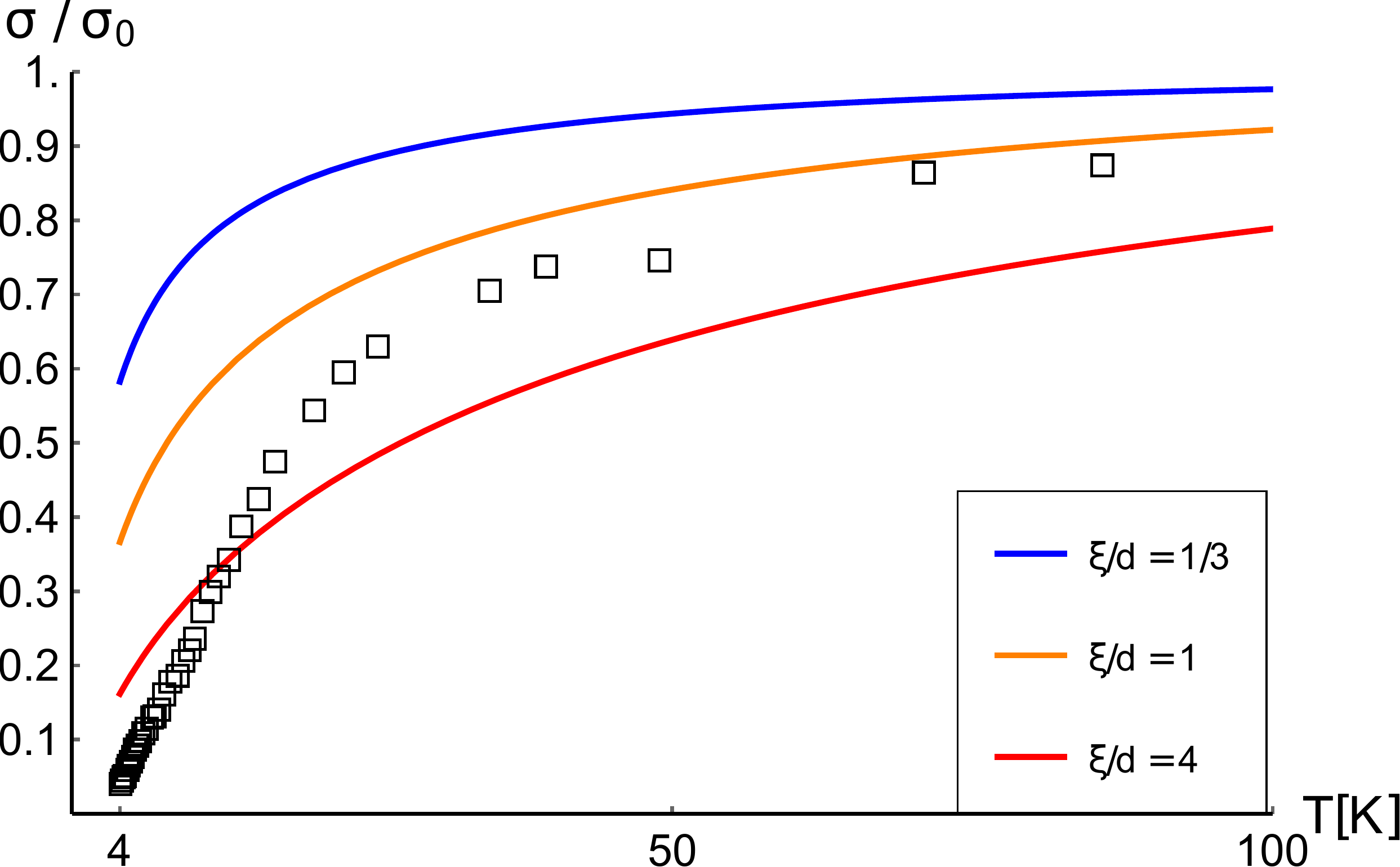}
\caption{\label{PolaronReduction} Polaronic reduction given by the ratio of the conductivity in the presence of the metal plate [Eq.~(\ref{CondSinglePh})] and the conductivity in the absence of the metal plate [Eq.~(\ref{MilleAbrahams})], plotted as function of temperature for $k_Fl = 4$, $\kappa = 1$ in Eq.~(\ref{JBal+Diff}) (see discussion in the last paragraph of the section) and different values of $\xi/d$ ($1/3,1$ and $4$ for upper, middle, and lower curves respectively) where $\xi$ is the localization length and $d$ is the EG-Metal separation. The blank square are taken from experiment~\cite{ZviCommunication}. The ratio is calculated for, $\tilde{\omega}=3K$, $d=10\text{nm}$, $\omega^D_c \approx 1500K$ (for Au layer) and $\kappa = 3$ for optimal hopping length [Eq.~(\ref{Optimal-ES-r})], values which are also compatible with the estimated parameters in Ref.~\cite{PhysRevLett.56.643}. As expected, increasing the value of $\xi/d$ leads to a larger value of optimal hopping length [Eq.~(\ref{Optimal-ES-r})] and consequently to a larger polaronic reduction [Eq.~(\ref{PofE-EH-Ohmic})]. The exprimental data of Ref.~\cite{PhysRevLett.56.643} is two orders of magnitude smaller than the data presented in squares, therefore it is presented only in logarithmic scale as indicated by circles in Fig.~\ref{PolaronReductionFit} below. Note that $\sigma$ includes only the dynamical effect of the metal plate. The static effect of the metal plate, within the ES regime given in Eq.~(\ref{DominancePolaron}), is small and therefore neglected.}
\end{figure}

The polaronic reduction in the macroscopic conductivity as a function of temperature is then given by substituting Eq.~(\ref{Optimal-ES-r}) in the conductivities with and without the metal layer [Eq.~(\ref{CondSinglePh}), Eq.~(\ref{MilleAbrahams}) respectivly). The resulted ratio between the conductivities with and without the metal as a function of temperature is presented in Fig.~\ref{PolaronReduction}, for parameters compatible with Ref.~\cite{PhysRevLett.56.643} see caption. As can be seen the polaron causes a reduction in the conductivity which becomes more appreciable at low temperatures. The blank circles represents the unpublished data ~\cite{ZviCommunication}. Our results show an appreciable reduction, yet a much weaker temperature dependence compared to the experimental data. 

\begin{figure}
\includegraphics[scale=0.32]{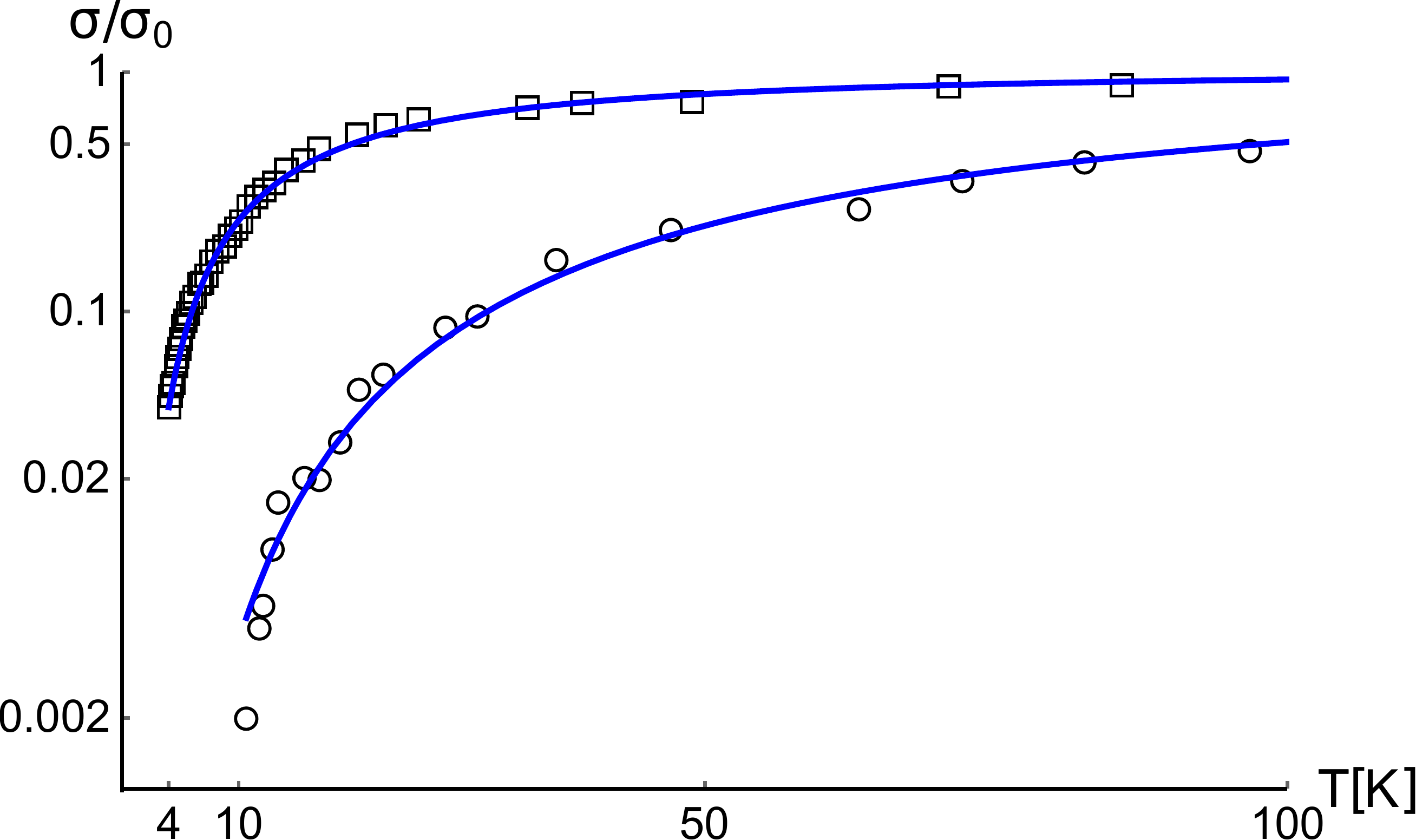}
\caption{\label{PolaronReductionFit} 
Log plot of the ratio of the conductivities with and without the metal layer as a function of temperature. The experiments are given by squares~\cite{ZviCommunication} with Au metal layer and circles~\cite{PhysRevLett.56.643} with Ag metal layer. In both cases $\omega^D_c = 1500K$, $\kappa = 3$, $d = 10nm$ and $\tilde{\omega} = 3K$. The solid lines are the theory given by the ratio of Eq.~(\ref{CondSinglePh}) and Eq.~(\ref{MilleAbrahams}). The fit to squares is given with $\xi/d = 0.22$, with a fitting multiplication factor to Eq.~(\ref{DEHCoupling}) of $\frac{C}{k_F l\kappa^2} = 2.3$. The fit to circles with $\xi/d = 1.3$, with fit parameter $\frac{C}{k_F l\kappa^2}= 2$.}
\end{figure}

We note that our results have strong sensitivity to the magnitude of the screening of the EG-Metal interaction. Reduction in the screening induces an increase in the effective interaction, and with it a stronger polaon effect and stronger reduction of the conductivity at low temperatures. In Fig.~\ref{PolaronReductionFit} we fit the experimental data allowing enlarged effective interaction. This allows very good fits to experiments with other parameter values being compatible with experimental values. Investigating the origin of such an increase in the effective interaction between the EG and the metallic layer is beyond the scope of this paper. Yet, possible mechanisms are scattering sources in the metal which effectively reduce the screening or plasma modes which serve as an additional source of dissipation.

As can be seen from Eq.~(\ref{DEHCoupling}) and Eq.~(\ref{PofE-EH-Ohmic}) the polaronic effect is sensitive to the sample parameters such as scattering length and dielectric constant. This is consistent with the large differences in the conductivity observed for different samples \cite{PhysRevLett.56.643,ZviCommunication} [see also Fig.~\ref{PolaronReductionFit}]. Furthermore, an additional source for the said differences in the samples can be disorder in the EG layer. One can see that for larger disorder in the EG layer (lower curve in Fig.~\ref{PolaronReductionFit}) there is a stronger reduction in the conductivity (caused by the presence of the metal layer). This may originate from the screening within the EG layer, allowed by its finite width, which decreases for larger disorder and consequently increases the EG-Metal interaction. Thus a larger disorder may allow for a stronger EG-Metal interaction.


In the case where Mott VRH dominates the low temperature regime (i.e. for $T_3 \sim T_1$), one should replace Eq.~(\ref{Optimal-ES-r}) by the optimal hopping distance compatible with constant DOS ($g_0$). In this regime we find qualitatively similar behaviour as shown in Fig.~\ref{PolaronReduction} for the same temperature range.

Throughout the paper we neglected spatial variation and frequency dependence of the dielectric constant ($\kappa$). Taking such variations into account will have only a quantitative effect on our results, which can be negated by making corresponding changes to other unknown parameters in the system. In general, we note that for the dynamical (polaronic) response [given in Eq.~(\ref{JBal+Diff})] we expect dielectric constant values to be smaller than static dielectric constant values, in accordance with fitting values chosen in our calculations.

\subsection{Microscopic explanation of the polaronic reduction}
\label{MicroscopicOrigin}

\begin{figure}
\includegraphics[scale=0.28]{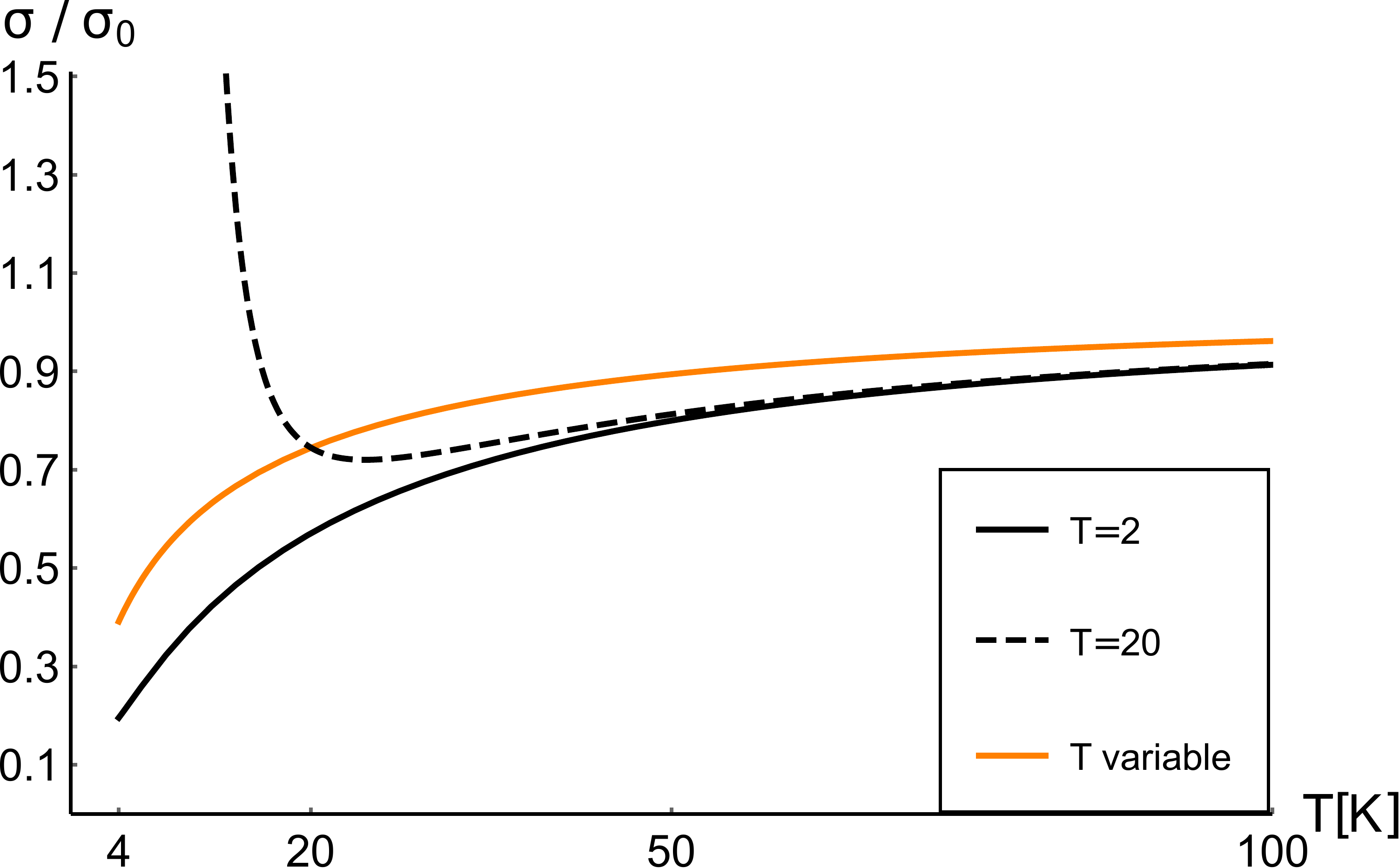}
\caption{\label{ElectronAssist}
Ratio of the EG conductivity in the presence and in the absence of a metal layer [Eq.~(\ref{CondSinglePh}) and Eq.~(\ref{MilleAbrahams})] is plotted as function of phonon temperature. The orange curve (upper solid curve) has the same parameters as the orange curve plotted in Fig.~\ref{PolaronReduction} and describes the situation of metallic plate having the same temperature as that of phonons. To demonstrate the contribution of the metal electrons to the phonon assisted tunneling in the EG we plot the above ratio of conductivities keeping the temperature of the metal constant at 2K represented by the solid black curve (lower solid curve) and at 20K (dashed black line).}
\end{figure}

In this paper we discuss the effect of the electrons in the metallic plate on the phonon assisted tunneling in the EG. This is not the only effect of the metallic plate. As Eq.~(\ref{Convolution}) suggests, the 
electrons of the metallic plate can also assist hopping in the EG by providing the thermal energy and can even 
make hopping possible in absence of phonons. Our results show that this is a subdominant effect as long as the temperatures of the metallic plate and that of phonons are equal. Thus, in total, the metallic plate causes an overall suppression of the conductivity. In Fig.~\ref{ElectronAssist} we present a detailed analysis attesting to the competing effect of the metal electrons on the phonon assisted tunneling in the EG. Keeping the metal electrons at a constant low temperature (lower than the phonon temperature in the whole relevant domain), one make the suppression by the metallic plate even stronger. However, if the temperature of the metal electrons becomes higher than that of the phonons, we find an enhanced conductance, in accordance with the above picture.

Furthermore, comparing between the processes in which the electronic and phononic environments assist the EG electron at the same temperature [i.e. comparing $P^{el}(E<0)$ and $\tilde{\omega}^{-2} |E| n^{ph}(|E|)$ respectively, see Eq.~(\ref{CondSinglePh})], one can show that for a given energy the electronic assistance is much smaller than the phononic assistance. This can be explained by the small ratio of the prefactors of each process - $\tilde{\omega}/\omega^D_c \ll 1$ [where $\tilde{\omega}$ and $\omega^D_c$ are defined in and Eq.~(\ref{El-PhEnergyScale}) and Eq.~(\ref{DiffusionCutoffFerquency}) respectively]. Thus, we expect that for $\tilde{\omega}/\omega^D_c \sim 1$ the electronic environment would give rise to an overall increase in the conductance. One should note though that this condition might not exist in real systems.

\section{Conclusions}
\label{Conclusions}

In this work we studied the polaronic effect on the conductivity of a two dimensional electron-glass system coupled to phonons and in proximity to a metal layer. The metal layer effectively screens the Coulomb interactions, and also dresses the electron's tunneling amplitude in the EG system. The latter is also known as the polaronic effect. Using field theoretical approach we have derived an effective action for the system and obtained an expression for the conductivity to leading order in the dressed tunneling amplitude. Since the disorder is the largest energy scale in the EG system the approximated conductivity is valid for a wide range of coupling strengths to the phonon and electron environments. We further approximated the conductivity retaining only the single phonon process and found that the polaronic effect causes, in a wide temperature regime, a reduction in the VRH conductivity by up to an order of magnitude. The main mechanism of the dynamical polaronic effect is that extra activation energy 
must be provided by the phonons assisted tunneling process in the EG, thus reducing its probability. We also found that the logarithmic dependence of the polaronic reduction on distance does not change the exponent $p$ in both ES's and Mott's VRH regimes. Our results are in agreement with the overall trend in experiment~\cite{PhysRevLett.56.643, ZviCommunication}. However, to obtain good quantitative fit with experiment we must assume an effectively larger coupling constant. This may originate from an additional contribution of plasma modes, dynamical response of the insulator, or a reduced screening caused by additional scattering mechanisms in the metal. Furthermore, the conductivity varies greatly between different samples~\cite{ZviCommunication}. This can be the result of the polaronic effect being sensitive to a small changes in the mean free path, dielectric constant, and the effective screening of both layers.

\section{Acknowledgments}

We would like to thank Zvi Ovadyahu for illuminating discussions that initiated this project and followed it throughout. We would also like to thank Igor Gornyi, Alexander Mirlin, Peter Nalbach, Dmitri Polyakov and the late Joe Imry for useful discussions.

The work of AB is partially supported by the US National Science Foundation (CHE-1462075), NSF and Louisiana Board of Regents LINK Program and Carrol Lavin Bernick Foundation Research Grant (2020).
MS acknowledges support from the Israel Science Foundation (Grant No. 821/14 and Grant No. 2300/19).

\appendix

\section{Solution of the Mean-field equation}
\label{MFE-Sol}

To solve the MF equation we expand the generalized propagator [Eq.~(\ref{GeneralizedGreenFunc})] in powers of $\Delta\Phi^0$ [defined in Eq.~(\ref{ansatz2})]:
\begin{equation}
\label{GeneralizedPropagaorExpansion}
\begin{split}
&2i\sum_{k,\omega}\left(G_{V,\Phi^0}\right)_{(k \; \omega),(k + q \;\omega + \Omega)}\\
&= 2i\sum_{k,\omega}G_{0k+q,\omega+\Omega} \sum_n \la k,\omega|\left(G_0 i\Delta\bar{\Phi}^0\right)^n|k + q,\omega + \Omega\ra \\ 
&\approx 2i\sum_{k,\omega}G_{0k+q,\omega+\Omega} \la k,\omega|\sum_{\substack{k',\omega' \\ q',\Omega'}}G_{0k'\omega'}\left(i\Delta\Phi^0_{q',\Omega'}\right)|k',\omega'\ra \times \\
&\quad\quad\quad\quad \times \la k'+q',\omega' + \Omega'|k + q,\omega + \Omega\ra \\
&= -2\sum_{k,\omega}G_{0k\omega}G_{0k+q,\omega+\Omega}\,\Delta\Phi^0_{q,\Omega} \equiv -\Pi_{q\Omega}\,\Delta\Phi^0_{q,\Omega}\, ,
\end{split}
\end{equation}
where in the last equality we used Eq.~(\ref{Polarization}). The zeroth order contribution cancels because of electro-neutrality.
Substituting Eq.~(\ref{GeneralizedPropagaorExpansion}) in the MF equation [Eq.~(\ref{Mean-Field})] we get:
\begin{equation}
\label{ApproMFSmallFluc}
\begin{split}
0 &= \left(V^{(2)}_q\right)^{-1} \Phi^0_{q\Omega} -\Pi_{q\Omega} \Delta\Phi^0_{q,\Omega}.
\end{split}
\end{equation}
Solving then for $\Phi^0$ we obtain Eq.~(\ref{MFE-ApproxSol}).

\section{Gaussian corrections to MF solution}
\label{Fluctuations}

Representing the action in terms of the deviation from the saddle point solution, $\delta\Phi$, we get:
\begin{equation}
\label{S_OmegaHubbardDeltaPhi2}
\begin{split}
&\mathcal{S} = \mathcal{S}_\text{EG} + \mathcal{S}_t + \frac{1}{2} \sum_{q,\Omega} \delta\bar{\Phi}_{q,\Omega} \left(V^{(2)}_q\right)^{-1} \delta\Phi_{q,\Omega}\\
&\quad - \frac{1}{2} \sum_{i,j,\Omega, q} (1-f_{q\Omega})^2\left(V^{(2)}_q\right)^{-1} V^{(1)*}_{qi}V^{(1)}_{qj}\bar{n}_{i,\Omega}n_{j,\Omega}\\
&\quad + i\sum_{j,\Omega,q}(1-f_{q\Omega})\left(V^{(2)}_q\right)^{-1} V^{(1)}_{qj}n_{j,\Omega}\delta\bar{\Phi}_{q,\Omega}\\
&\quad - \text{Tr\,ln}\left(-G^{-1}_0 + i\Delta\Phi^0 + i\delta\bar{\Phi}\right)\, ,
\end{split}
\end{equation}
where $\Delta\Phi^0$ is given in Eq.~(\ref{ansatz2}).

The leading terms coming from the fluctuations around the MF solution, $\Phi^0$, can be found by expanding Eq.~(\ref{S_OmegaHubbardDeltaPhi2}) to second order in $\delta\Phi$: 
\begin{equation}
\label{SeffMF+fluctuations1}
\begin{split}
\mathcal{S}_\text{FL} &= \mathcal{S}_\text{EG} + \frac{1}{2} \sum_{i,j,\Omega} \tilde{K}_{ij}\bar{n}_{i,\Omega}n_{j,\Omega} + \text{Tr}\left[\sum_{n\geq 3} \frac{F^n}{n}\right]\\
&\quad - \frac{1}{2}\Tr \left[\left(\mathbb{1} - F\right)^{-1}G_0\delta\bar{\Phi}\left(\mathbb{1} - F\right)^{-1}G_0\delta\bar{\Phi}\right]\\
&\quad + \frac{1}{2}\sum_{\Omega, q}\delta\Phi_{q,\Omega} \left(V^{(2)}_q\right)^{-1}\delta\Phi_{q,\Omega}\, ,
\end{split}
\end{equation}
where $F = iG_0\Delta\bar{\Phi}^0$, and the constant $\text{Tr\,ln}\left(-G^{-1}_0\right)$ is disregarded [see discussion below Eq.~(\ref{S_MF2})]. Note that given our MF solution is approximate, a linear term should be finite, however since we consider $(q_{TF}d)^{-1}\ll 1$ [see Eq.~(\ref{Deviation})] it is neglected. Furthermore, from now on we keep track only on corrections to the two-body terms of the EG degrees of freedom, assuming that the three-body and higher terms are negligible. This is typical in localized systems where the disorder is larger than the average near neighbour interactions.
Denoting the first line in Eq.~(\ref{SeffMF+fluctuations1}) as $\mathcal{S}_\text{MF}$ [see Eq.~(\ref{S_MF3})], and integrating over $\delta\Phi$ fields we obtain:
\begin{equation}
\label{SeffMF+fluctuations2}
\begin{split}
&\mathcal{S}_\text{FL} = \mathcal{S}_\text{MF} + \text{Tr\,ln}A\, ,
\end{split}
\end{equation}
with,
\begin{equation}
\label{DefA}
A_{qq'\Omega\Omega'} = \left(V^{(2)}_q\right)^{-1}\delta_{q,q'}\delta_{\Omega,\Omega'} - M_{qq'\omega\omega'}\, ,
\end{equation}
where
\begin{equation}
\begin{split}
\label{DefM}
M_{qq'\Omega\Omega'} &= \sum_{\substack{n,m \\ k,\omega\\ k_1,\omega_1}} G_{0k_1\omega_1} G_{0k+q',\omega+\Omega'}\times \\
&\qquad\;\; \times(F^n)_{kk_1\omega\omega_1} (F^m)_{k_1+q,k+q',\omega_1+\Omega,\omega+\Omega'}\\
&\approx M^{(0)}_{qq'\Omega\Omega'} + M^{(1)}_{qq'\Omega\Omega'} + M^{(2)}_{qq'\Omega\Omega'}.
\end{split}
\end{equation}
Here $M^{(i)}$ is the $i$'th order term:
\begin{equation}
\begin{split}
\label{M-Orders}
M_{qq'\Omega\Omega'}^{(0)} &= \Pi_{q\Omega}\delta_{qq'}\delta_{\Omega\Omega'}\\
M_{qq'\Omega\Omega'}^{(1)} &= \sum_{i,\omega, k}\mathcal{G}^{(1)}_{qq'\Omega\Omega'} f_{q-q',\Omega-\Omega'} V^{(1)}_{q-q',i}n_{i,\Omega-\Omega'}
\\
M_{qq'\Omega\Omega'}^{(2)} &= \sum_{i,j, k, \omega} \mathcal{G}^{(2)}_{qq'\Omega\Omega'} f_{q_1,\Omega_1}f_{q-q'-q_1,\Omega-\Omega'-\Omega_1} \\
&\quad\quad\quad \times V^{(1)}_{q_1,i}V^{(1)}_{q-q'-q_1,j} n_{i,\Omega_1} n_{j,\Omega-\Omega'-\Omega_1}\\
&\quad + \sum_{i,j, k, \omega} \mathcal{G}^{(3)}_{qq'\Omega\Omega'} f_{q+q_1,\Omega+\Omega_1}f_{q'+q_1,\Omega'+\Omega_1} \\
&\quad\quad\quad \times V^{(1)}_{q+q_1,i} V^{(1)*}_{q'+q_1,j}n_{i,\Omega+\Omega_1}\bar{n}_{j,\Omega'+\Omega_1}\, ,
\end{split}
\end{equation}
with
\begin{equation}
\begin{split}
\mathcal{G}^{(1)}_{qq'\Omega\Omega'} &= \left(G_{0k+q'-q,\omega+\Omega'-\Omega} + G_{0k+q,\omega+\Omega}\right) \\
&\qquad\; \times G_{0k\omega} G_{0k+q',\omega+\Omega'}\\
\mathcal{G}^{(2)}_{qq'\Omega\Omega'} &= \sum_{q_1,\Omega_1}\left(G_{0k+q-q_1,\omega+\Omega-\Omega_1} + G_{0k+q'-q,\omega+\Omega'-\Omega}\right)\\
&\quad\quad\quad \times G_{0k\omega}G_{0k+q',\omega+\Omega'}G_{0k-q_1,\omega-\Omega_1}\\
\mathcal{G}^{(3)}_{qq'\Omega\Omega'} &= \sum_{q_1,\Omega_1} G_{0k\omega} G_{0k+q',\omega+\Omega'} G_{0k+q,\omega+\Omega}G_{0k-k_1,\omega-\Omega_1}.
\end{split}
\end{equation}
Substituting Eq.~(\ref{DefA}) and Eq.~(\ref{DefM}) back into Eq.~(\ref{SeffMF+fluctuations2}) we find:
\begin{equation}
\label{SeffMF+fluctuations2ndOrderInF}
\begin{split}
&\delta \mathcal{S}=\mathcal{S}_\text{FL} - \mathcal{S}_\text{MF} \approx \text{Tr\,ln}\left[\left(V^{(2)RPA}\right)^{-1}-M^{(1)}-M^{(2)}\right]\, ,
\end{split}
\end{equation}
where we denote the corrections to the MF action as $\delta \mathcal{S}$.
Further expanding Eq.~(\ref{SeffMF+fluctuations2ndOrderInF}) to order $Tr(F^2)$ we obtain:
\begin{equation}
\label{SeffMF+fluctuations2ndOrderInF2}
\begin{split}
\delta\mathcal{S} &\approx - \text{Tr\,ln}\left(V^{(2)RPA}\right) - \sum_{\Omega,q}V^{(2)RPA}_{q\Omega}M^{(2)}_{q\Omega}\\
&\quad - \frac{1}{2}\sum_{\Omega,q,\Omega',q'}\left(V^{(2)RPA}_{q\Omega}\right)^2 M^{(1)}_{qq'\Omega\Omega'}M^{(1)}_{q'q\Omega'\Omega}\, ,
\end{split}
\end{equation}
where $V^{(2)RPA}_{q\Omega} = f_{q\Omega}V^{(2)}_q$ is the screened interaction in the metal layer in the RPA approximation. Comparing the general form of the correction terms to the kernel obtained from the MF (0th order) contribution, one can see that the 1st and 2nd corrections in Eq.~(\ref{SeffMF+fluctuations2ndOrderInF2}) goes as $\sim G^4 f^3 V^{(2)}(V^{(1)})^2$, $\sim G^6 f^4 (V^{(2)})^2(V^{(1)})^2$, respectively, while the MF kernel is $\tilde{K}\sim G^2 f(V^{(1)})^2$. Since we consider the regime where $1/q_{TF}d \ll 1$ we can roughly approximate $f_q \lesssim 1/q_{TF}d$ [see Eq.~(\ref{ProofSmallDeltaPhi})], therefore the MF term (which has the smallest power in $f$) is larger than the fluctuation corrections. Furthermore, as can be seen from Eq.~(\ref{M-Orders}), the $M^{(1)}$ and $M^{(2)}$ terms involve high order correlations of the free electron propagator $G_0$. Roughly speaking, a larger correlation means integration over effectively smaller phase space than less correlated terms, such as the RPA.

\section{The generating functional and the conductivity}
\label{GeneratingFunctionalApp}

In order to obtain the matrix of conductances $\sigma_{ij}$ we define a generating functional by adding a source term to the effective action [given in Eq.~(\ref{ModelAction})] as: $\mathcal{S}_U = i\int_0^{1} d\tau \sum_i U_i(\tau)n_i(\tau)$ where $U_i(\tau)$ is the classical external potential field at site $i$ in the EG system. Performing a gauge transformation to eliminate $\mathcal{S}_U$ amounts to a shift of the gauge field given in Eq.~(\ref{ModelAction}), $\Theta_{ij}(\tau) \rightarrow \Theta_{ij}(\tau) + \chi_{ij}(\tau)$:
\begin{equation}
\label{GeneratingAction}
\begin{aligned}
\mathcal{S}_\text{eff}[\chi] &= \tilde{\mathcal{S}}_\text{EG} + \int_0^{1} d\tau \sum_{i\neq j} \bar{c}_i(\tau) \tilde{t}_{ij}e^{i\chi_{ij}(\tau)} c_j(\tau) \\
&\; + \frac{1}{2}\int_0^{1} \int_0^{1} d\tau d\tau' \sum_{i,j}\phi_i(\tau) \left(K'\right)^{-1}_{ij}(\tau-\tau') \phi_j(\tau')\, ,
\end{aligned}
\end{equation}
where $\chi_{ij}=\chi_i-\chi_j$ and $\chi_i(\tau) = \int_0^{\tau}d\tau' \,U_i(\tau')$.
Starting from the current,
\begin{equation}
\label{Current}
\begin{split}
&\la I_i(\tau)[\chi]\ra = i\frac{\delta}{\delta \chi_i(\tau)} \ln \mathcal{Z}[\chi]\\
& = \sum_{j(\neq i)} \left\langle \left(\bar{c}_i(\tau) \tilde{t}_{i j}e^{i\chi_{i j\tau}}c_j(\tau) - \bar{c}_j(\tau) \tilde{t}_{i j}^{*}e^{-i\chi_{i j\tau}}c_i(\tau)\right)\right\rangle\, ,
\end{split}
\end{equation}
and expanding to linear order in $U$ we get:
\begin{equation}
\label{ConductivityDef}
\la I_{i\Omega}[\chi]\ra \approx - \frac{1}{\Omega}\sum_j C^I_{ij\Omega} U_{j,\Omega}\, ,
\end{equation}
where $C^I_{ij\Omega} = \int_0^{1}d\tau e^{i\Omega \tau}C^I_{ij}(\tau)$ is the response function  defined by the generating functional, $\mathcal{Z}[\chi]$,
\begin{equation}
\begin{split}
\label{ConductivityKernel}
C^I_{ij}(\tau-\tau') = \frac{1}{\mathcal{Z}}\frac{\delta^2}{\delta \chi_j(\tau')\delta \chi_i(\tau)}\mathcal{Z}[\chi]\Big|_{\chi=0}.
\end{split}
\end{equation}
Here the average is defined as,
\begin{equation}
\label{Average}
\la \mathcal{O}\ra = \frac{1}{\mathcal{Z}[\chi]}\int \mathcal{D}[\bar{c},c,\phi] \mathcal{O} e^{-\mathcal{S}_\text{eff}[\chi]}\, ,
\end{equation}
where $\mathcal{O}$ is some functional operator and the generating functional is:
\begin{equation}
\label{GeneratingFunctional}
\mathcal{Z}[\chi] = \int \mathcal{D}[\bar{c},c,\phi]e^{-\mathcal{S}_\text{eff}[\chi]}\, ,
\end{equation}
where $\mathcal{S}_\text{eff}[\chi]$ is defined in Eq.~(\ref{GeneratingAction}). Performing then the variational derivatives in Eq.~(\ref{ConductivityKernel}) gives Eq.~(\ref{CondKernelExplicit}).

The DC conductance between sites $i$ and $j$ is obtained by the following steps: (1) we calculate the response function $C^I_{ij\Omega}$ according to Eq.~(\ref{ConductivityKernel}), (2) expand $C^I_{ij\Omega}$ to leading order in the dressed tunnelling amplitude, (3) perform the analytical continuation to real time and then to real frequency, (4) take the static limit and, (5) represent Eq.~(\ref{ConductivityDef}) in terms of potential drop between sites $i$ and $j$, $U_{ij}$. The DC conductance matrix $\sigma_{ij}$ is then defined as follows:
\begin{equation}
\begin{split}
\label{Conductivity}
\la I_i\ra &\approx -\lim_{\omega\rightarrow 0}\frac{i}{\beta\omega}\left[\sum_j C^I_{ij\Omega}U_{j,\Omega}\right]\Big|_{\Omega \rightarrow -i\beta\omega + \delta} \\
&=-\lim_{\omega\rightarrow 0}\frac{i}{\beta\omega}\sum_j C^I_{ij\omega}U_{j,\omega} = \sum_j \sigma_{ij} U_{ij}\ .
\end{split}
\end{equation}
The analytical continuation to real time (frequency $\omega$) of Eq.~(\ref{ConductivityKernel}) is implemented by the prescription given in Ref.~\cite{Efetov2003}. The last equality in Eq.~(\ref{Conductivity}) is the definition of the conductance matrix\cite{Shklovskii1984,PhysRevB.80.245214}; this form is obtained by substituting the expression given in Eq.~(\ref{FinalCondKernel}) for the response function $C^I_{ij\omega}$.

\section{The correlation function of the gauge field}
\label{J-Evaluation}

Using Eq.~(\ref{TotalKernel}) and Eq.~(\ref{ModelAction}) we calculate the correlation function of the gauge field, $\Theta_{ij}$ with respect to the free action of the potential field $\phi$:
\begin{equation}
\begin{split}
\label{J-Correlation}
J_{ij}(\tau) &= \la\Theta_{ij}(\tau)\Theta_{ij}(0)\ra_{0\phi}-\la \Theta_{ij}(0)^2 \ra_{0\phi}\\
&= \sum_{\Omega\neq 0} \frac{e^{-i\Omega\tau}-1}{\Omega^2}\left(K'_{ii\Omega}+K'_{jj\Omega}-K'_{ij\Omega}-K'_{ji\Omega}\right)\\
& \equiv J^{el}_{ij}(\tau) + J^{ph}_{ij}(\tau)\, ,
\end{split}
\end{equation}
where $J^{el}_{ij}(\tau)$, $J^{ph}_{ij}(\tau)$ are respectively the e-h and phonon contributions which are given by:
\begin{equation}
\begin{split}
\label{J-Correlation2}
J^{el}_{ij}(\tau) &= 4\sum_{\substack{k,q \\ \Omega\neq 0}}\left(V^{(1)}_{qij}\right)^2 f_{q0}f_{q\Omega} \frac{e^{-i\Omega\tau}-1}{(i\Omega)^2}(\Pi_{q\Omega}-\Pi_{q0})\\
J^{ph}(\tau) &= 4\sum_{\substack{q, \Omega\neq 0}} |g_q|^2 \frac{e^{-i\Omega\tau}-1}{\omega_q(\omega_q^2 + \Omega^2)}\, ,
\end{split}
\end{equation}
with $V^{(1)}_{qij} = V^{(1)}_q\sin\left(\frac{\bm{qr}_{ij}}{2}\right)$ and $\Pi_{q\Omega}$ given in Eq.~(\ref{Polarization}). For the phonon correlation function we take the limit $qr_{ij} \gg 1$ which gives $\sin^2\left(\frac{\bm{qr}_{ij}}{2}\right) \approx 1/2$. This is valid for $r_{ij}/\xi \gg 1$ where $\xi$ is the localization length~\cite{MillerAbrahams1960,Shklovskii1984}. This is since the $r_{ij}$ dependent term of the phonon mediated interaction is short ranged and decays sufficiently fast for the typical near neighbour distance. Since our main interest is the polaron induced by the metal layer we keep the distance dependence of the e-h correlation function, $J^{el}$.
To perform the summation over $\Omega$ in Eq.~(\ref{J-Correlation2}) we represent it as an integral over the complex frequency plane with a contour that excludes the poles of the integrand and also the point $\Omega = 0$. The exclusion of $\Omega = 0$ gives an additional residue at the point $\Omega = 0$ in $J_{el}$ (and $J_{ph}$), which turns out to have exactly zero value.
Performing the summation over $\Omega$ in Eq.~(\ref{J-Correlation2}) and introducing the e-h and phonon spectral functions ($S^{el}(\omega)$, $S^{ph}(\omega)$ respectively) we obtain:
\begin{equation}
\begin{split}
\label{J-CorrLongWaveLimit}
J_{ij}^{el}(\tau) &= \int_0^{\infty} \frac{d\omega}{\omega^2}S^{el}_{ij}(\omega) F(\omega,\tau),\\
J^{ph}(\tau) &= \int_0^{\infty} \frac{d\omega}{\omega^2}S^{ph}(\omega)F(\omega,\tau)\, ,
\end{split}
\end{equation}
where,
\begin{equation}
\label{BosonicPart}
F(\omega,\tau) = \coth\left(\frac{\beta\omega}{2}\right)(\cosh(\omega \tau)-1) - \sinh(\omega \tau).
\end{equation}
As usual the phonon spectral function in the deformation potential approximation is super Ohmic~\cite{MillerAbrahams1960}:
\begin{equation}
\label{PhSpectralFunction}
S^{ph}(\omega)  \equiv \sum_q |g_q|^2 \delta(\omega-\omega_q) \approx \frac{\omega^s}{\tilde{\omega}^{s-1}}\Theta(\omega^{ph}_{c} - \omega)\, ,
\end{equation}
where $s$ is the spatial dimension. For $s=3$, 
\begin{equation}
\label{El-PhEnergyScale}
\tilde{\omega}^2 = \frac{2\pi \hbar^3 \rho c^5}{3\gamma^2},
\end{equation}
$c$ is speed of sound, $\rho$ is the mass density and $\gamma$ is the deformation potential averaged over the transverse and longitudinal directions. Furthermore, the form factor of the el-ph interaction has a powerlaw cutoff $(1+(\omega/\omega^{ph}_c)^2)^{-3}$ with the cutoff frequency $\omega^{ph}_c = 2c/\xi$~\cite{MillerAbrahams1960}. Since we are interested in the low energy behaviour we approximate the phonon form factor with a step cutoff.

The e-h spectral function is given by the imaginary part of the retarded electronic Kernel $K^{el}$ [given in Eq.~(\ref{ScreenedKernel})]. Using Eq.~(\ref{J-Correlation2}) and the definition in Eq.~(\ref{J-CorrLongWaveLimit}) we obtain:
\begin{equation}
\label{ElSpectralFunction}
\begin{split}
&S^{el}_{ij}(\omega) \equiv\\
&\equiv -\frac{1}{\pi} \lim_{\eta \rightarrow 0} \text{Im}\left[4\sum_q \left(V^{(1)}_{qij}\right)^2 f_{q0}f_{q,\omega+i\eta} (\Pi_{q,\omega+i\eta} - \Pi_{q0})\right]\\
&= -\frac{4}{\pi}\lim_{\eta \rightarrow 0} |f_{q,\omega + i\eta}|^2 \left(V^{(1)}_{qij}\right)^2 \text{Im}[\Pi_{q,\omega+i\eta}]\\
&= 4\sum_{k, q}\left(\lim_{\eta \rightarrow 0}|f_{q,\omega+i\eta}|^2\right) \left(V_{qij}^{(1)}\right)^2 N_{k+q,k} \delta (\omega-E_{k,k+q}) \\
& \approx \frac{1}{4 \pi \kappa^2}\frac{1}{k_F d}\omega G_{ij}(\omega/\omega_c)\, ,
\end{split}
\end{equation}  
where we perform the analytical continuation of the metal's kernel $K^{el}_{ij\Omega}\rightarrow K^{el}_{ij,\omega+i\eta}$, $N_{k+q,k}$ is defined below Eq.~(\ref{polarizationDisorder}) and we used the long wavelength limit [Eq.~(\ref{LongWaveLengthLimit})]. The form factor is:
\begin{equation}
\begin{split}
\label{Gfunc}
G_{ij}(x_1)=\int_0^{\infty}dx \frac{e^{-2\sqrt{x^2+x_1^2}}}{\sqrt{x^2+x_1^2}}\left[1-J_0\left(\sqrt{x^2+x_1^2} \, R_{ij}\right)\right].
\end{split}
\end{equation}
where $x = qd$, $R_{ij} = r_{ij}/d$, $J_0$ is the Bessel function, $x_1 = \omega/\omega_c$ with the cutoff frequency $\omega_c = 2E_F/k_Fd$ and the inequality $(q_{TF}d) \gg 1$ is used.
As can be deduced from Eq.~(\ref{Deviation}) the factor $f_{q\omega}$ diverges at the plasma frequency. Thus in the approximation given in Eq.~(\ref{ElSpectralFunction}) we assume that the plasma modes do not overlap with the e-h modes and therefore we approximate $f_{q,\omega+i\eta}$ as constant in the relevant regime ($0<q<1/d$), i.e. we set $\lim_{\eta \rightarrow 0}|f_{q,\omega+i\eta}|^2 \approx |f_{q,\omega}|^2 \approx f^2_{q,0}$.

The low frequency behaviour of the e-h spectral function is valid for
\begin{equation}
\label{ScalingLimit}
x_1 R_{ij}  = \frac{\omega}{\omega_c}\frac{r_{ij}}{d} \ll  1\, ,
\end{equation}
which gives an Ohmic spectral function in the zeroth order ($x_1 = 0$). Together with an approximated exponential cutoff [see Eq.~(\ref{Gfunc}) for $x_1 > 1$ values] we can approximate a simple form for the spectral function:
\begin{equation}
\label{ElOhmicSpectralFunction}
S^{el}_{ij}(\omega) \approx \alpha_{ij}\omega e^{-\omega/\omega_c}.
\end{equation}
The resulted dimensionless coupling constant is:
\begin{equation}
\label{EHCouplingApp}
\alpha_{ij} \approx \frac{1}{4\pi \kappa^2}\frac{1}{k_F d} G_{ij}(0)
\end{equation}
where,
\begin{equation}
G_{ij}(0) = \ln\left[\frac{1}{2}+\frac{1}{2}\sqrt{1+\left(\frac{r_{ij}}{2d}\right)^2}\right].
\end{equation}

The correlation functions of the e-h and phonon baths do not include the static contribution by definition [see Eq.~(\ref{J-Correlation})], i.e. $J^{el/ph}(\tau=0)=0$. The static contribution is known as the Debye-Waller exponent which takes the form:
\begin{equation}
\label{DWexponent}
W = \int_0^{\infty} \frac{d\omega}{\omega^2}S(\omega)\coth\left(\frac{\beta \omega}{2}\right).
\end{equation}
For an Ohmic bath, $S(\omega) \propto \omega$, $W$ has an infrared divergence which is responsible for the overall convergence of $J^{el}(\tau)$. However, for the super-Ohmic phonon bath $S(\omega) \propto \omega^s$ with $s>2$, $W$ has a finite value as in the case of phonons in three dimensions. 

\nocite{*}
\bibliography{EOIMOEG}

\begin{thebibliography}{40}%
\makeatletter
\providecommand \@ifxundefined [1]{%
 \@ifx{#1\undefined}
}%
\providecommand \@ifnum [1]{%
 \ifnum #1\expandafter \@firstoftwo
 \else \expandafter \@secondoftwo
 \fi
}%
\providecommand \@ifx [1]{%
 \ifx #1\expandafter \@firstoftwo
 \else \expandafter \@secondoftwo
 \fi
}%
\providecommand \natexlab [1]{#1}%
\providecommand \enquote  [1]{``#1''}%
\providecommand \bibnamefont  [1]{#1}%
\providecommand \bibfnamefont [1]{#1}%
\providecommand \citenamefont [1]{#1}%
\providecommand \href@noop [0]{\@secondoftwo}%
\providecommand \href [0]{\begingroup \@sanitize@url \@href}%
\providecommand \@href[1]{\@@startlink{#1}\@@href}%
\providecommand \@@href[1]{\endgroup#1\@@endlink}%
\providecommand \@sanitize@url [0]{\catcode `\\12\catcode `\$12\catcode
  `\&12\catcode `\#12\catcode `\^12\catcode `\_12\catcode `\%12\relax}%
\providecommand \@@startlink[1]{}%
\providecommand \@@endlink[0]{}%
\providecommand \url  [0]{\begingroup\@sanitize@url \@url }%
\providecommand \@url [1]{\endgroup\@href {#1}{\urlprefix }}%
\providecommand \urlprefix  [0]{URL }%
\providecommand \Eprint [0]{\href }%
\providecommand \doibase [0]{http://dx.doi.org/}%
\providecommand \selectlanguage [0]{\@gobble}%
\providecommand \bibinfo  [0]{\@secondoftwo}%
\providecommand \bibfield  [0]{\@secondoftwo}%
\providecommand \translation [1]{[#1]}%
\providecommand \BibitemOpen [0]{}%
\providecommand \bibitemStop [0]{}%
\providecommand \bibitemNoStop [0]{.\EOS\space}%
\providecommand \EOS [0]{\spacefactor3000\relax}%
\providecommand \BibitemShut  [1]{\csname bibitem#1\endcsname}%
\let\auto@bib@innerbib\@empty
\bibitem [{\citenamefont {Anderson}(1958)}]{PhysRev.109.1492}%
  \BibitemOpen
  \bibfield  {author} {\bibinfo {author} {\bibfnamefont {P.~W.}\ \bibnamefont
  {Anderson}},\ }\href {\doibase 10.1103/PhysRev.109.1492} {\bibfield
  {journal} {\bibinfo  {journal} {Phys. Rev.}\ }\textbf {\bibinfo {volume}
  {109}},\ \bibinfo {pages} {1492} (\bibinfo {year} {1958})}\BibitemShut
  {NoStop}%
\bibitem [{\citenamefont {N.F.Mott}(1969)}]{N.F.Mott}%
  \BibitemOpen
  \bibfield  {author} {\bibinfo {author} {\bibnamefont {N.F.Mott}},\ }\href
  {\doibase 10.1080/14786436908216338} {\bibfield  {journal} {\bibinfo
  {journal} {The Philosophical Magazine: A Journal of Theoretical Experimental
  and Applied Physics}\ }\textbf {\bibinfo {volume} {19}},\ \bibinfo {pages}
  {835} (\bibinfo {year} {1969})},\ \Eprint
  {http://arxiv.org/abs/https://doi.org/10.1080/14786436908216338}
  {https://doi.org/10.1080/14786436908216338} \BibitemShut {NoStop}%
\bibitem [{\citenamefont {Efros}\ and\ \citenamefont
  {Shklovskii}(1975)}]{Efros_1975}%
  \BibitemOpen
  \bibfield  {author} {\bibinfo {author} {\bibfnamefont {A.~L.}\ \bibnamefont
  {Efros}}\ and\ \bibinfo {author} {\bibfnamefont {B.~I.}\ \bibnamefont
  {Shklovskii}},\ }\href {\doibase 10.1088/0022-3719/8/4/003} {\ \textbf
  {\bibinfo {volume} {8}},\ \bibinfo {pages} {L49} (\bibinfo {year}
  {1975})}\BibitemShut {NoStop}%
\bibitem [{\citenamefont {Shafarman}\ \emph {et~al.}(1989)\citenamefont
  {Shafarman}, \citenamefont {Koon},\ and\ \citenamefont
  {Castner}}]{PhysRevB.40.1216}%
  \BibitemOpen
  \bibfield  {author} {\bibinfo {author} {\bibfnamefont {W.~N.}\ \bibnamefont
  {Shafarman}}, \bibinfo {author} {\bibfnamefont {D.~W.}\ \bibnamefont {Koon}},
  \ and\ \bibinfo {author} {\bibfnamefont {T.~G.}\ \bibnamefont {Castner}},\
  }\href {\doibase 10.1103/PhysRevB.40.1216} {\bibfield  {journal} {\bibinfo
  {journal} {Phys. Rev. B}\ }\textbf {\bibinfo {volume} {40}},\ \bibinfo
  {pages} {1216} (\bibinfo {year} {1989})}\BibitemShut {NoStop}%
\bibitem [{\citenamefont {Rosenbaum}(1991)}]{PhysRevB.44.3599}%
  \BibitemOpen
  \bibfield  {author} {\bibinfo {author} {\bibfnamefont {R.}~\bibnamefont
  {Rosenbaum}},\ }\href {\doibase 10.1103/PhysRevB.44.3599} {\bibfield
  {journal} {\bibinfo  {journal} {Phys. Rev. B}\ }\textbf {\bibinfo {volume}
  {44}},\ \bibinfo {pages} {3599} (\bibinfo {year} {1991})}\BibitemShut
  {NoStop}%
\bibitem [{\citenamefont {Sarachik}\ and\ \citenamefont
  {Dai}(2002)}]{Sarachik_2002}%
  \BibitemOpen
  \bibfield  {author} {\bibinfo {author} {\bibfnamefont {M.~P.}\ \bibnamefont
  {Sarachik}}\ and\ \bibinfo {author} {\bibfnamefont {P.}~\bibnamefont {Dai}},\
  }\href {\doibase 10.1209/epl/i2002-00164-y} {\bibfield  {journal} {\bibinfo
  {journal} {Europhysics Letters ({EPL})}\ }\textbf {\bibinfo {volume} {59}},\
  \bibinfo {pages} {100} (\bibinfo {year} {2002})}\BibitemShut {NoStop}%
\bibitem [{\citenamefont {Amir}\ \emph {et~al.}(2009)\citenamefont {Amir},
  \citenamefont {Oreg},\ and\ \citenamefont {Imry}}]{PhysRevB.80.245214}%
  \BibitemOpen
  \bibfield  {author} {\bibinfo {author} {\bibfnamefont {A.}~\bibnamefont
  {Amir}}, \bibinfo {author} {\bibfnamefont {Y.}~\bibnamefont {Oreg}}, \ and\
  \bibinfo {author} {\bibfnamefont {Y.}~\bibnamefont {Imry}},\ }\href {\doibase
  10.1103/PhysRevB.80.245214} {\bibfield  {journal} {\bibinfo  {journal} {Phys.
  Rev. B}\ }\textbf {\bibinfo {volume} {80}},\ \bibinfo {pages} {245214}
  (\bibinfo {year} {2009})}\BibitemShut {NoStop}%
\bibitem [{\citenamefont {Entin-Wohlman}\ and\ \citenamefont
  {Ovadyahu}(1986)}]{PhysRevLett.56.643}%
  \BibitemOpen
  \bibfield  {author} {\bibinfo {author} {\bibfnamefont {O.}~\bibnamefont
  {Entin-Wohlman}}\ and\ \bibinfo {author} {\bibfnamefont {Z.}~\bibnamefont
  {Ovadyahu}},\ }\href {\doibase 10.1103/PhysRevLett.56.643} {\bibfield
  {journal} {\bibinfo  {journal} {Phys. Rev. Lett.}\ }\textbf {\bibinfo
  {volume} {56}},\ \bibinfo {pages} {643} (\bibinfo {year} {1986})}\BibitemShut
  {NoStop}%
\bibitem [{\citenamefont {Hu}\ \emph {et~al.}(1995)\citenamefont {Hu},
  \citenamefont {Keuls}, \citenamefont {Carmi}, \citenamefont {Jiang},\ and\
  \citenamefont {Dahm}}]{HU199565}%
  \BibitemOpen
  \bibfield  {author} {\bibinfo {author} {\bibfnamefont {X.}~\bibnamefont
  {Hu}}, \bibinfo {author} {\bibfnamefont {F.~V.}\ \bibnamefont {Keuls}},
  \bibinfo {author} {\bibfnamefont {Y.}~\bibnamefont {Carmi}}, \bibinfo
  {author} {\bibfnamefont {H.}~\bibnamefont {Jiang}}, \ and\ \bibinfo {author}
  {\bibfnamefont {A.}~\bibnamefont {Dahm}},\ }\href {\doibase
  https://doi.org/10.1016/0038-1098(95)00401-7} {\bibfield  {journal} {\bibinfo
   {journal} {Solid State Communications}\ }\textbf {\bibinfo {volume} {96}},\
  \bibinfo {pages} {65 } (\bibinfo {year} {1995})}\BibitemShut {NoStop}%
\bibitem [{\citenamefont {van Keuls}\ \emph {et~al.}(1996)\citenamefont {van
  Keuls}, \citenamefont {Hu}, \citenamefont {Dahm},\ and\ \citenamefont
  {W.}}]{VANKEULS1996945}%
  \BibitemOpen
  \bibfield  {author} {\bibinfo {author} {\bibfnamefont {F.~W.}\ \bibnamefont
  {van Keuls}}, \bibinfo {author} {\bibfnamefont {X.~L.}\ \bibnamefont {Hu}},
  \bibinfo {author} {\bibfnamefont {A.}~\bibnamefont {Dahm}}, \ and\ \bibinfo
  {author} {\bibfnamefont {J.~H.}\ \bibnamefont {W.}},\ }\href {\doibase
  https://doi.org/10.1016/0039-6028(96)00570-5"} {\bibfield  {journal}
  {\bibinfo  {journal} {Surface Science}\ }\textbf {\bibinfo {volume}
  {361-362}},\ \bibinfo {pages} {945 } (\bibinfo {year} {1996})}\BibitemShut
  {NoStop}%
\bibitem [{\citenamefont {Adkins}\ and\ \citenamefont
  {Astrakharchik}(1998)}]{Adkins_1998}%
  \BibitemOpen
  \bibfield  {author} {\bibinfo {author} {\bibfnamefont {C.~J.}\ \bibnamefont
  {Adkins}}\ and\ \bibinfo {author} {\bibfnamefont {E.~G.}\ \bibnamefont
  {Astrakharchik}},\ }\href {\doibase 10.1088/0953-8984/10/30/006} {\bibfield
  {journal} {\bibinfo  {journal} {Journal of Physics: Condensed Matter}\
  }\textbf {\bibinfo {volume} {10}},\ \bibinfo {pages} {6651} (\bibinfo {year}
  {1998})}\BibitemShut {NoStop}%
\bibitem [{\citenamefont {Bennaceur}\ \emph {et~al.}(2012)\citenamefont
  {Bennaceur}, \citenamefont {Jacques}, \citenamefont {Portier}, \citenamefont
  {Roche},\ and\ \citenamefont {Glattli}}]{EStoMott3}%
  \BibitemOpen
  \bibfield  {author} {\bibinfo {author} {\bibfnamefont {K.}~\bibnamefont
  {Bennaceur}}, \bibinfo {author} {\bibfnamefont {P.}~\bibnamefont {Jacques}},
  \bibinfo {author} {\bibfnamefont {F.}~\bibnamefont {Portier}}, \bibinfo
  {author} {\bibfnamefont {P.}~\bibnamefont {Roche}}, \ and\ \bibinfo {author}
  {\bibfnamefont {D.~C.}\ \bibnamefont {Glattli}},\ }\href {\doibase
  10.1103/PhysRevB.86.085433} {\bibfield  {journal} {\bibinfo  {journal} {Phys.
  Rev. B}\ }\textbf {\bibinfo {volume} {86}},\ \bibinfo {pages} {085433}
  (\bibinfo {year} {2012})}\BibitemShut {NoStop}%
\bibitem [{\citenamefont {Ovadyahu}(2019)}]{PhysRevB.99.184201}%
  \BibitemOpen
  \bibfield  {author} {\bibinfo {author} {\bibfnamefont {Z.}~\bibnamefont
  {Ovadyahu}},\ }\href {\doibase 10.1103/PhysRevB.99.184201} {\bibfield
  {journal} {\bibinfo  {journal} {Phys. Rev. B}\ }\textbf {\bibinfo {volume}
  {99}},\ \bibinfo {pages} {184201} (\bibinfo {year} {2019})}\BibitemShut
  {NoStop}%
\bibitem [{\citenamefont {Yakimov}\ \emph
  {et~al.}(2000{\natexlab{a}})\citenamefont {Yakimov}, \citenamefont
  {Dvurechenskii}, \citenamefont {Nikiforov},\ and\ \citenamefont
  {Adkins}}]{EStoMottQD1}%
  \BibitemOpen
  \bibfield  {author} {\bibinfo {author} {\bibfnamefont {A.}~\bibnamefont
  {Yakimov}}, \bibinfo {author} {\bibfnamefont {A.}~\bibnamefont
  {Dvurechenskii}}, \bibinfo {author} {\bibfnamefont {A.}~\bibnamefont
  {Nikiforov}}, \ and\ \bibinfo {author} {\bibfnamefont {C.}~\bibnamefont
  {Adkins}},\ }\href {\doibase
  10.1002/(SICI)1521-3951(200003)218:1<99::AID-PSSB99>3.0.CO;2-7} {\bibfield
  {journal} {\bibinfo  {journal} {physica status solidi (b)}\ }\textbf
  {\bibinfo {volume} {218}},\ \bibinfo {pages} {99} (\bibinfo {year}
  {2000}{\natexlab{a}})}\BibitemShut {NoStop}%
\bibitem [{\citenamefont {Yakimov}\ \emph
  {et~al.}(2000{\natexlab{b}})\citenamefont {Yakimov}, \citenamefont
  {Dvurechenskii}, \citenamefont {Kirienko}, \citenamefont {Yakovlev},
  \citenamefont {Nikiforov},\ and\ \citenamefont {Adkins}}]{EStoMottQD2}%
  \BibitemOpen
  \bibfield  {author} {\bibinfo {author} {\bibfnamefont {A.~I.}\ \bibnamefont
  {Yakimov}}, \bibinfo {author} {\bibfnamefont {A.~V.}\ \bibnamefont
  {Dvurechenskii}}, \bibinfo {author} {\bibfnamefont {V.~V.}\ \bibnamefont
  {Kirienko}}, \bibinfo {author} {\bibfnamefont {Y.~I.}\ \bibnamefont
  {Yakovlev}}, \bibinfo {author} {\bibfnamefont {A.~I.}\ \bibnamefont
  {Nikiforov}}, \ and\ \bibinfo {author} {\bibfnamefont {C.~J.}\ \bibnamefont
  {Adkins}},\ }\href {\doibase 10.1103/PhysRevB.61.10868} {\bibfield  {journal}
  {\bibinfo  {journal} {Phys. Rev. B}\ }\textbf {\bibinfo {volume} {61}},\
  \bibinfo {pages} {10868} (\bibinfo {year} {2000}{\natexlab{b}})}\BibitemShut
  {NoStop}%
\bibitem [{\citenamefont {Aleiner}\ and\ \citenamefont
  {Shklovskii}(1994)}]{PhysRevB.49.13721}%
  \BibitemOpen
  \bibfield  {author} {\bibinfo {author} {\bibfnamefont {I.~L.}\ \bibnamefont
  {Aleiner}}\ and\ \bibinfo {author} {\bibfnamefont {B.~I.}\ \bibnamefont
  {Shklovskii}},\ }\href {\doibase 10.1103/PhysRevB.49.13721} {\bibfield
  {journal} {\bibinfo  {journal} {Phys. Rev. B}\ }\textbf {\bibinfo {volume}
  {49}},\ \bibinfo {pages} {13721} (\bibinfo {year} {1994})}\BibitemShut
  {NoStop}%
\bibitem [{\citenamefont {Larkin}\ and\ \citenamefont
  {Khmel’nitskii}(1982)}]{larkin1982activation}%
  \BibitemOpen
  \bibfield  {author} {\bibinfo {author} {\bibfnamefont {A.}~\bibnamefont
  {Larkin}}\ and\ \bibinfo {author} {\bibfnamefont {D.}~\bibnamefont
  {Khmel’nitskii}},\ }\href
  {http://www.jetp.ac.ru/cgi-bin/e/index/e/56/3/p647?a=list} {\bibfield
  {journal} {\bibinfo  {journal} {Sov. Phys. JETP}\ }\textbf {\bibinfo {volume}
  {56}},\ \bibinfo {pages} {647} (\bibinfo {year} {1982})}\BibitemShut
  {NoStop}%
\bibitem [{\citenamefont {Miller}\ and\ \citenamefont
  {Abrahams}(1960)}]{MillerAbrahams1960}%
  \BibitemOpen
  \bibfield  {author} {\bibinfo {author} {\bibfnamefont {A.}~\bibnamefont
  {Miller}}\ and\ \bibinfo {author} {\bibfnamefont {E.}~\bibnamefont
  {Abrahams}},\ }\href {\doibase 10.1103/PhysRev.120.745} {\bibfield  {journal}
  {\bibinfo  {journal} {Phys. Rev.}\ }\textbf {\bibinfo {volume} {120}},\
  \bibinfo {pages} {745} (\bibinfo {year} {1960})}\BibitemShut {NoStop}%
\bibitem [{\citenamefont {Altland}\ and\ \citenamefont
  {Simons}(2010)}]{altland_simons_2010}%
  \BibitemOpen
  \bibfield  {author} {\bibinfo {author} {\bibfnamefont {A.}~\bibnamefont
  {Altland}}\ and\ \bibinfo {author} {\bibfnamefont {B.~D.}\ \bibnamefont
  {Simons}},\ }\href {\doibase 10.1017/CBO9780511789984} {\emph {\bibinfo
  {title} {Condensed Matter Field Theory}}},\ \bibinfo {edition} {2nd}\ ed.\
  (\bibinfo  {publisher} {Cambridge University Press},\ \bibinfo {year}
  {2010})\BibitemShut {NoStop}%
\bibitem [{\citenamefont {Tsvelik}(2003)}]{Tsvelik_2003}%
  \BibitemOpen
  \bibfield  {author} {\bibinfo {author} {\bibfnamefont {A.~M.}\ \bibnamefont
  {Tsvelik}},\ }\href {\doibase 10.1017/CBO9780511615832} {\emph {\bibinfo
  {title} {Quantum Field Theory in Condensed Matter Physics}}},\ \bibinfo
  {edition} {2nd}\ ed.\ (\bibinfo  {publisher} {Cambridge University Press},\
  \bibinfo {year} {2003})\BibitemShut {NoStop}%
\bibitem [{\citenamefont {Fradkin}(2013)}]{fradkin_2013}%
  \BibitemOpen
  \bibfield  {author} {\bibinfo {author} {\bibfnamefont {E.}~\bibnamefont
  {Fradkin}},\ }\href {\doibase 10.1017/CBO9781139015509} {\emph {\bibinfo
  {title} {Field Theories of Condensed Matter Physics}}},\ \bibinfo {edition}
  {2nd}\ ed.\ (\bibinfo  {publisher} {Cambridge University Press},\ \bibinfo
  {year} {2013})\BibitemShut {NoStop}%
\bibitem [{\citenamefont {Pikus}\ and\ \citenamefont
  {Efros}(1995)}]{PhysRevB.51.16871}%
  \BibitemOpen
  \bibfield  {author} {\bibinfo {author} {\bibfnamefont {F.~G.}\ \bibnamefont
  {Pikus}}\ and\ \bibinfo {author} {\bibfnamefont {A.~L.}\ \bibnamefont
  {Efros}},\ }\href {\doibase 10.1103/PhysRevB.51.16871} {\bibfield  {journal}
  {\bibinfo  {journal} {Phys. Rev. B}\ }\textbf {\bibinfo {volume} {51}},\
  \bibinfo {pages} {16871} (\bibinfo {year} {1995})}\BibitemShut {NoStop}%
\bibitem [{\citenamefont {Ingold}\ and\ \citenamefont
  {Nazarov}(1992)}]{NazarovIngold1992}%
  \BibitemOpen
  \bibfield  {author} {\bibinfo {author} {\bibfnamefont {G.-L.}\ \bibnamefont
  {Ingold}}\ and\ \bibinfo {author} {\bibfnamefont {Y.~V.}\ \bibnamefont
  {Nazarov}},\ }\enquote {\bibinfo {title} {Charge tunneling rates in
  ultrasmall junctions},}\ in\ \href {\doibase 10.1007/978-1-4757-2166-9_2}
  {\emph {\bibinfo {booktitle} {Single Charge Tunneling: Coulomb Blockade
  Phenomena In Nanostructures}}},\ \bibinfo {editor} {edited by\ \bibinfo
  {editor} {\bibfnamefont {H.}~\bibnamefont {Grabert}}\ and\ \bibinfo {editor}
  {\bibfnamefont {M.~H.}\ \bibnamefont {Devoret}}}\ (\bibinfo  {publisher}
  {Springer US},\ \bibinfo {address} {Boston, MA},\ \bibinfo {year} {1992})\
  pp.\ \bibinfo {pages} {21--107}\BibitemShut {NoStop}%
\bibitem [{\citenamefont {Efetov}\ and\ \citenamefont
  {Tschersich}(2003)}]{Efetov2003}%
  \BibitemOpen
  \bibfield  {author} {\bibinfo {author} {\bibfnamefont {K.~B.}\ \bibnamefont
  {Efetov}}\ and\ \bibinfo {author} {\bibfnamefont {A.}~\bibnamefont
  {Tschersich}},\ }\href {\doibase 10.1103/PhysRevB.67.174205} {\bibfield
  {journal} {\bibinfo  {journal} {Phys. Rev. B}\ }\textbf {\bibinfo {volume}
  {67}},\ \bibinfo {pages} {174205} (\bibinfo {year} {2003})}\BibitemShut
  {NoStop}%
\bibitem [{\citenamefont {Shklovskii}\ and\ \citenamefont
  {Efros}(1984)}]{Shklovskii1984}%
  \BibitemOpen
  \bibfield  {author} {\bibinfo {author} {\bibfnamefont {B.~I.}\ \bibnamefont
  {Shklovskii}}\ and\ \bibinfo {author} {\bibfnamefont {A.~L.}\ \bibnamefont
  {Efros}},\ }\enquote {\bibinfo {title} {A general description of hopping
  conduction in lightly doped semiconductors},}\ in\ \href {\doibase
  10.1007/978-3-662-02403-4_4} {\emph {\bibinfo {booktitle} {Electronic
  Properties of Doped Semiconductors}}}\ (\bibinfo  {publisher} {Springer
  Berlin Heidelberg},\ \bibinfo {address} {Berlin, Heidelberg},\ \bibinfo
  {year} {1984})\ pp.\ \bibinfo {pages} {74--93}\BibitemShut {NoStop}%
\bibitem [{\citenamefont {L\"u}\ and\ \citenamefont
  {Zheng}(2007)}]{VarPolaron1}%
  \BibitemOpen
  \bibfield  {author} {\bibinfo {author} {\bibfnamefont {Z.}~\bibnamefont
  {L\"u}}\ and\ \bibinfo {author} {\bibfnamefont {H.}~\bibnamefont {Zheng}},\
  }\href {\doibase 10.1103/PhysRevB.75.054302} {\bibfield  {journal} {\bibinfo
  {journal} {Phys. Rev. B}\ }\textbf {\bibinfo {volume} {75}},\ \bibinfo
  {pages} {054302} (\bibinfo {year} {2007})}\BibitemShut {NoStop}%
\bibitem [{\citenamefont {Zhao}\ \emph {et~al.}(2011)\citenamefont {Zhao},
  \citenamefont {L\"u},\ and\ \citenamefont {Zheng}}]{VarPolaron2}%
  \BibitemOpen
  \bibfield  {author} {\bibinfo {author} {\bibfnamefont {C.}~\bibnamefont
  {Zhao}}, \bibinfo {author} {\bibfnamefont {Z.}~\bibnamefont {L\"u}}, \ and\
  \bibinfo {author} {\bibfnamefont {H.}~\bibnamefont {Zheng}},\ }\href
  {\doibase 10.1103/PhysRevE.84.011114} {\bibfield  {journal} {\bibinfo
  {journal} {Phys. Rev. E}\ }\textbf {\bibinfo {volume} {84}},\ \bibinfo
  {pages} {011114} (\bibinfo {year} {2011})}\BibitemShut {NoStop}%
\bibitem [{\citenamefont {Nazir}\ \emph {et~al.}(2012)\citenamefont {Nazir},
  \citenamefont {McCutcheon},\ and\ \citenamefont {Chin}}]{VarPolaron3}%
  \BibitemOpen
  \bibfield  {author} {\bibinfo {author} {\bibfnamefont {A.}~\bibnamefont
  {Nazir}}, \bibinfo {author} {\bibfnamefont {D.~P.~S.}\ \bibnamefont
  {McCutcheon}}, \ and\ \bibinfo {author} {\bibfnamefont {A.~W.}\ \bibnamefont
  {Chin}},\ }\href {\doibase 10.1103/PhysRevB.85.224301} {\bibfield  {journal}
  {\bibinfo  {journal} {Phys. Rev. B}\ }\textbf {\bibinfo {volume} {85}},\
  \bibinfo {pages} {224301} (\bibinfo {year} {2012})}\BibitemShut {NoStop}%
\bibitem [{\citenamefont {Lee}\ and\ \citenamefont
  {Ramakrishnan}(1985)}]{RevModPhys.57.287}%
  \BibitemOpen
  \bibfield  {author} {\bibinfo {author} {\bibfnamefont {P.~A.}\ \bibnamefont
  {Lee}}\ and\ \bibinfo {author} {\bibfnamefont {T.~V.}\ \bibnamefont
  {Ramakrishnan}},\ }\href {\doibase 10.1103/RevModPhys.57.287} {\bibfield
  {journal} {\bibinfo  {journal} {Rev. Mod. Phys.}\ }\textbf {\bibinfo {volume}
  {57}},\ \bibinfo {pages} {287} (\bibinfo {year} {1985})}\BibitemShut
  {NoStop}%
\bibitem [{\citenamefont {Belitz}\ and\ \citenamefont
  {Kirkpatrick}(1994)}]{RevModPhys.66.261}%
  \BibitemOpen
  \bibfield  {author} {\bibinfo {author} {\bibfnamefont {D.}~\bibnamefont
  {Belitz}}\ and\ \bibinfo {author} {\bibfnamefont {T.~R.}\ \bibnamefont
  {Kirkpatrick}},\ }\href {\doibase 10.1103/RevModPhys.66.261} {\bibfield
  {journal} {\bibinfo  {journal} {Rev. Mod. Phys.}\ }\textbf {\bibinfo {volume}
  {66}},\ \bibinfo {pages} {261} (\bibinfo {year} {1994})}\BibitemShut
  {NoStop}%
\bibitem [{\citenamefont {Kamenev}(2011)}]{kamenev_2011}%
  \BibitemOpen
  \bibfield  {author} {\bibinfo {author} {\bibfnamefont {A.}~\bibnamefont
  {Kamenev}},\ }\href {\doibase 10.1017/CBO9781139003667} {\emph {\bibinfo
  {title} {Field Theory of Non-Equilibrium Systems}}}\ (\bibinfo  {publisher}
  {Cambridge University Press},\ \bibinfo {year} {2011})\BibitemShut {NoStop}%
\bibitem [{\citenamefont {Zala}\ \emph {et~al.}(2001)\citenamefont {Zala},
  \citenamefont {Narozhny},\ and\ \citenamefont
  {Aleiner}}]{BallisticDiffusiveCrossover2001}%
  \BibitemOpen
  \bibfield  {author} {\bibinfo {author} {\bibfnamefont {G.}~\bibnamefont
  {Zala}}, \bibinfo {author} {\bibfnamefont {B.~N.}\ \bibnamefont {Narozhny}},
  \ and\ \bibinfo {author} {\bibfnamefont {I.~L.}\ \bibnamefont {Aleiner}},\
  }\href {\doibase 10.1103/PhysRevB.64.214204} {\bibfield  {journal} {\bibinfo
  {journal} {Phys. Rev. B}\ }\textbf {\bibinfo {volume} {64}},\ \bibinfo
  {pages} {214204} (\bibinfo {year} {2001})}\BibitemShut {NoStop}%
\bibitem [{\citenamefont {Leggett}\ \emph {et~al.}(1987)\citenamefont
  {Leggett}, \citenamefont {Chakravarty}, \citenamefont {Dorsey}, \citenamefont
  {Fisher}, \citenamefont {Garg},\ and\ \citenamefont {Zwerger}}]{Leggett1987}%
  \BibitemOpen
  \bibfield  {author} {\bibinfo {author} {\bibfnamefont {A.~J.}\ \bibnamefont
  {Leggett}}, \bibinfo {author} {\bibfnamefont {S.}~\bibnamefont
  {Chakravarty}}, \bibinfo {author} {\bibfnamefont {A.~T.}\ \bibnamefont
  {Dorsey}}, \bibinfo {author} {\bibfnamefont {M.~P.~A.}\ \bibnamefont
  {Fisher}}, \bibinfo {author} {\bibfnamefont {A.}~\bibnamefont {Garg}}, \ and\
  \bibinfo {author} {\bibfnamefont {W.}~\bibnamefont {Zwerger}},\ }\href
  {\doibase 10.1103/RevModPhys.59.1} {\bibfield  {journal} {\bibinfo  {journal}
  {Rev. Mod. Phys.}\ }\textbf {\bibinfo {volume} {59}},\ \bibinfo {pages} {1}
  (\bibinfo {year} {1987})}\BibitemShut {NoStop}%
\bibitem [{\citenamefont {Kagan}\ and\ \citenamefont
  {Prokof'ev}(1987)}]{kagan1987quantum}%
  \BibitemOpen
  \bibfield  {author} {\bibinfo {author} {\bibfnamefont {Y.}~\bibnamefont
  {Kagan}}\ and\ \bibinfo {author} {\bibfnamefont {N.}~\bibnamefont
  {Prokof'ev}},\ }\href@noop {} {\bibfield  {journal} {\bibinfo  {journal} {Zh.
  Eksp. Teor. Fiz.}\ }\textbf {\bibinfo {volume} {93}},\ \bibinfo {pages} {366}
  (\bibinfo {year} {1987})}\BibitemShut {NoStop}%
\bibitem [{\citenamefont {B\"uttiker}\ and\ \citenamefont
  {Landauer}(1982)}]{ButtikerLandauer}%
  \BibitemOpen
  \bibfield  {author} {\bibinfo {author} {\bibfnamefont {M.}~\bibnamefont
  {B\"uttiker}}\ and\ \bibinfo {author} {\bibfnamefont {R.}~\bibnamefont
  {Landauer}},\ }\href {\doibase 10.1103/PhysRevLett.49.1739} {\bibfield
  {journal} {\bibinfo  {journal} {Phys. Rev. Lett.}\ }\textbf {\bibinfo
  {volume} {49}},\ \bibinfo {pages} {1739} (\bibinfo {year}
  {1982})}\BibitemShut {NoStop}%
\bibitem [{\citenamefont {Bulka}\ \emph {et~al.}(1985)\citenamefont {Bulka},
  \citenamefont {Kramer},\ and\ \citenamefont {MacKinnon}}]{WlargerThenFermi1}%
  \BibitemOpen
  \bibfield  {author} {\bibinfo {author} {\bibfnamefont {B.}~\bibnamefont
  {Bulka}}, \bibinfo {author} {\bibfnamefont {B.}~\bibnamefont {Kramer}}, \
  and\ \bibinfo {author} {\bibfnamefont {A.}~\bibnamefont {MacKinnon}},\
  }\href@noop {} {\bibfield  {journal} {\bibinfo  {journal} {Zeitschrift f\"ur
  Physik B Condensed Matter}\ }\textbf {\bibinfo {volume} {60}},\ \bibinfo
  {pages} {13} (\bibinfo {year} {1985})}\BibitemShut {NoStop}%
\bibitem [{\citenamefont {Bulka}\ \emph {et~al.}(1987)\citenamefont {Bulka},
  \citenamefont {Schreiber},\ and\ \citenamefont {Kramer}}]{WlargerThenFermi2}%
  \BibitemOpen
  \bibfield  {author} {\bibinfo {author} {\bibfnamefont {B.}~\bibnamefont
  {Bulka}}, \bibinfo {author} {\bibfnamefont {M.}~\bibnamefont {Schreiber}}, \
  and\ \bibinfo {author} {\bibfnamefont {B.}~\bibnamefont {Kramer}},\
  }\href@noop {} {\bibfield  {journal} {\bibinfo  {journal} {Zeitschrift f\"ur
  Physik B Condensed Matter}\ }\textbf {\bibinfo {volume} {66}},\ \bibinfo
  {pages} {21} (\bibinfo {year} {1987})}\BibitemShut {NoStop}%
\bibitem [{\citenamefont {Ovadyahu}(2017)}]{DisorderStrengthZvi}%
  \BibitemOpen
  \bibfield  {author} {\bibinfo {author} {\bibfnamefont {Z.}~\bibnamefont
  {Ovadyahu}},\ }\href {\doibase 10.1103/PhysRevB.95.134203} {\bibfield
  {journal} {\bibinfo  {journal} {Phys. Rev. B}\ }\textbf {\bibinfo {volume}
  {95}},\ \bibinfo {pages} {134203} (\bibinfo {year} {2017})}\BibitemShut
  {NoStop}%
\bibitem [{\citenamefont {Aharony}\ \emph {et~al.}(1992)\citenamefont
  {Aharony}, \citenamefont {Zhang},\ and\ \citenamefont
  {Sarachik}}]{AharonyES-MottCrossover}%
  \BibitemOpen
  \bibfield  {author} {\bibinfo {author} {\bibfnamefont {A.}~\bibnamefont
  {Aharony}}, \bibinfo {author} {\bibfnamefont {Y.}~\bibnamefont {Zhang}}, \
  and\ \bibinfo {author} {\bibfnamefont {M.~P.}\ \bibnamefont {Sarachik}},\
  }\href {\doibase 10.1103/PhysRevLett.68.3900} {\bibfield  {journal} {\bibinfo
   {journal} {Phys. Rev. Lett.}\ }\textbf {\bibinfo {volume} {68}},\ \bibinfo
  {pages} {3900} (\bibinfo {year} {1992})}\BibitemShut {NoStop}%
\bibitem [{Zvi()}]{ZviCommunication}%
  \BibitemOpen
  \href@noop {} {}\bibinfo {note} {Z. Ovadyahu, private
  communication.}\BibitemShut {Stop}%
\end{thebibliography}%
\end{document}